\documentclass[11pt]{article}

\usepackage{comment}
\usepackage[utf8]{inputenc}
\usepackage[T1]{fontenc}
\usepackage{lmodern}
\usepackage{authblk}
\usepackage{geometry}
\usepackage[numbers,sort&compress]{natbib} % for \citep
\usepackage{graphicx}
\usepackage{float}     % for [H]
\usepackage{subfig}    % for \subfloat
\usepackage{changepage}% for adjustwidth
\newlength{\extralength}
\usepackage[hidelinks]{hyperref}
\usepackage{orcidlink}

\title{Urban Buildings Energy Consumption Estimation Using HPC:\\
A Case Study of Bologna}

\author[1,2]{Aldo Canfora\,\orcidlink{0009-0006-2770-4413}\thanks{Corresponding author: aldo.canfora2@unibo.it}}
\author[2]{Eleonora Bergamaschi\,\orcidlink{0009-0007-5370-0261}}
\author[2]{Riccardo Mioli\,\orcidlink{0009-0006-5323-3053}}
\author[3]{Federico Battini\,\orcidlink{0000-0002-1237-6430}}
\author[1]{Mirko Degli Esposti\,\orcidlink{0000-0003-0316-3449}}
\author[2]{Giorgio Pedrazzi\,\orcidlink{0000-0001-8896-3044}}
\author[2]{Chiara Dellacasa\,\orcidlink{0000-0002-8123-1737}}

\affil[1]{Department of Physics and Astronomy, University of Bologna,\\
Via Irnerio 46, 40126 Bologna, Italy}
\affil[2]{HPC Department, Cineca,\\
Via Magnanelli 6/3, 40033 Casalecchio di Reno, Italy}
\affil[3]{IFAB Foundation, Via Galliera 32, 40121 Bologna, Italy}

\date{November 2025}

\begin{document}

\maketitle

\begin{abstract}
Urban Building Energy Modeling (UBEM) plays a central role in understanding and forecasting energy consumption at the city scale. In this work, we present a UBEM pipeline that integrates EnergyPlus simulations, high-performance computing (HPC), and open geospatial datasets to estimate the energy demand of buildings in Bologna, Italy. Geometric information---including building footprints and heights---was obtained from the Bologna Open Data portal and enhanced with aerial LiDAR measurements. Non-geometric attributes such as construction materials, insulation characteristics, and window performance were derived from regional building regulations and the European TABULA database. The computation was carried out on Leonardo, the Cineca-hosted supercomputer, enabling the simulation of approximately 25{,}000 buildings in under 30 minutes.
\end{abstract}

\textbf{Keywords:} HPC simulations; UBEM; Urban Digital Twin; EnergyPlus; Ray

\section{Introduction}

One of the most pressing issues in the fight against climate change is the energy demand and carbon footprint of buildings.
Globally, buildings are responsible for nearly 30\% of greenhouse gas emissions, making them a critical sector for targeted interventions \citep{buildings_IEA}.
Similar findings were also reported by Chen \textit{et al.} (2017) who noted that buildings in cities consume 30–70\% of total primary energy, making the retrofit of existing building stock a key strategy for mitigating global warming \citep{CHEN2017323}. 
In densely populated urban areas, understanding how much energy each building consumes, where consumption is highest, and how emissions are distributed across the city is fundamental to optimize both public and private efforts towards decarbonization.
Such knowledge is not only essential for managing resources and planning strategic interventions (e.g., renovations, retrofitting, energy system upgrades), but also for allocating public funding more effectively.

Urban Building Energy Modeling (UBEM) is an emerging field that aims to simulate and understand the energy performance of buildings at the neighborhood or city scale.
Unlike traditional single-building energy models, UBEMs leverage geospatial datasets and building typologies to produce bottom-up simulations of energy demand, $CO_2$ emissions, and retrofit scenarios across thousands of buildings \citep{Ferrando_2020, JOHARI2020109902}.

In their work, Hong \textit{et al.} (2020) emphasized that collective modeling of building stocks can improve energy efficiency, resilience, and sustainability. 
Moreover, the authors pointed out that UBEM tools integrated with high-performance computing and rich datasets are crucial for planning large-scale retrofits and achieving climate goals \citep{HONG2020106508}.

Numerous studies have explored the development of a urban energy model in the context of real cities, highlighting its important role in adaptation planning.
Many of these works focus on the potential benefits achievable through retrofitting scenarios, while others analyse the most influential drivers of buildings energy consumption.

One of the most influential contributions in the field has come from the MIT Sustainable Design Lab, led by Prof. Christoph Reinhart.
Their seminal work published in 2016, demonstrated a replicable workflow for constructing UBEMs using publicly available GIS data and archetype libraries. 
By simulating the energy profiles of over 83,000 buildings in Boston, the team showed how UBEM could be employed to evaluate citywide interventions, such as PV installations or district energy networks, and assess their technical and economic viability.
Their research highlights the importance of accurate archetype definition, model calibration, and integration with real-world data such as energy audits or metered consumption \citep{CEREZODAVILA2016237}. 

A similar work by Mutani \textit{et al.} (2020) applied a GIS-based model in Turin, revealing that older, poorly insulated buildings in the city center had the highest energy demand and that low-income areas faced greater energy poverty risks, highlighting the importance of linking energy modeling with social equity \citep{su12145678}.

Buckley \textit{et al.} (2021) applied UBEM to 9,000 residential buildings in Dublin, identifying retrofit strategies capable of reducing GHG emissions by up to 60\% by 2030, with insulation and airtightness improvements offering the highest cost-effectiveness \citep{BUCKLEY_2021}.

Ji \textit{et al.} (2022) developed a bottom-up UBEM for Wuhan and found that commercial and transport buildings exhibited the highest energy use intensity, dominated by cooling loads. 
Accounting for shading and topography helped improving accuracy in complex urban settings \citep{Ji_2022}.

Deng \textit{et al.} (2023) applied a calibrated UBEM to 483 residential buildings in Geneva, projecting that by 2050 heating demand will decrease by 22-31\% while cooling demand could rise by 95-173\%. 
They further showed that envelope retrofits could reduce total heating and cooling use by 41.7\% and 18.6\%, respectively, demonstrating UBEM's value in climate adaptation planning \citep{DENG_2023}.

Garcia-Nevado \textit{et al.} (2025) showed that adding open balconies to retrofitted housing in Barcelona slightly increased heating demand by $0.1-1.6 kWh/m^2 yr$, but significantly reduced cooling demand by $0.1-3.8 kWh/m^2 yr$, with combined insulation-and-balcony retrofits lowering block-level energy demand by up to 16\% \citep{urbansci9110439}.

Espino-Reyes \textit{et al.} (2025) developed a national reference model to assess multi-family housing in eight Mexican cities, revealing that domestic water heating is the dominant driver of variable energy consumption, while heating and cooling demands depend strongly on local climate. 
Their results showed close agreement between simulated and measured electricity use, highlighting the potential of reference buildings as tools for guiding low and nearly zero energy housing design in Mexico \citep{urbansci9040113}.

Various approaches exist for building energy modeling. Early models were based purely on physical principles, whereas more advanced approaches integrate statistical methods to incorporate phenomena that cannot be fully explained by physics alone.
Foucquier \textit{et al.} (2013) demonstrated that coupling detailed physical models (e.g., CFD) with simplified nodal methods improves accuracy while reducing computation time. 
Their analysis also found that hybrid grey-box models offer an optimal trade-off between efficiency and precision, making them ideal for large-scale urban energy modeling \citep{FOUCQUIER_2013}.

There is also a foundational review of the UBEM domain \citep{REINHART2016196}, distinguishing it from earlier top-down approaches and advocating for physical, bottom-up simulations that incorporate heat transfer dynamics, occupancy patterns, and microclimatic effects.

Ali \textit{et al.} (2021) analyzed over two decades of UBEM research and reported that bottom-up physics-based models dominate (2368 studies—nearly four times more than data-driven approaches), with about 70\% of applications focusing on energy analysis and optimization. 
The study emphasized the need for hybrid models that integrate technical precision with urban-scale planning \citep{ALI2021111073}.

Cerezo \textit{et al.} (2017) demonstrated that Bayesian-calibrated probabilistic models reduced Energy Use Intensity (EUI) prediction errors by up to 70\% compared to deterministic approaches, enhancing the reliability of UBEM predictions at district scale \citep{CEREZO2017321}.

Oraiopoulos and Howard (2022) reviewed validated UBEM studies and found that model accuracy varies widely with spatial and temporal resolution—ranging from aggregate annual errors below 1\% to individual building errors exceeding 1000\%. 
They showed that Bayesian calibration consistently improved hourly accuracy, underscoring the importance of transparent calibration practices \citep{ORAIOPOULOS_2022}.

More recent and advanced approaches incorporate machine learning techniques and data driven methods, demonstrating the potential to enhance model accuracy.
However, the reliability of these models strongly depends on the availability of high quality datasets.
This approach was applied by Kontokosta and Tull (2017) who developed a predictive model for New York City using machine learning to estimate electricity and natural gas consumption for over 1.1 million buildings, demonstrating that statistical models can generalize well beyond benchmarked properties. 
Their study highlights that reliable city-scale predictions of building energy consumption can support evidence-based energy policies, helping planners identify inefficiencies and target interventions to reduce carbon emissions \citep{KONTOKOSTA2017303}.

Todeschi \textit{et al.} (2021) compared a machine learning model and a GIS-based engineering model for space heating in Fribourg, achieving mean absolute percentage errors of 12.8\% and 19.3\%, respectively. 
Model performance varied with building age and geometry, supporting the use of hybrid UBEM approaches \citep{su13041595}.

Another representative study is that of Wei \textit{et al.} (2018), showing that data-driven methods, such as NNs and SVMs, achieve high accuracy in predicting and classifying building energy use. 
Hybrid models significantly improve precision and efficiency compared to traditional simulations \citep{WEI20181027}.

In their work, Blanco \textit{et al.} (2024) introduced a data-driven framework to classify urban areas into standardized Urban Energy Units (UEUs) using open data and machine learning. 
Their method achieved up to 84\% accuracy in predicting building age and type, demonstrating its value in improving spatial energy demand mapping and municipal heat planning \citep{BLANCO2024105075}.

Many tools were developed to assist with the construction of an UBEM.
Depending on the scope and the kind of data available, different choiches can be made.

Ferrando \textit{et al.} (2020) compared major bottom-up UBEM tools and found that while these models provide detailed spatiotemporal simulations, they differ in accuracy, usability, and computational demand. 
The lack of standardized formats and validation benchmarks remains a key barrier to widespread adoption \citep{FERRANDO2020102408}.

Malhotra \textit{et al.} (2022) examined 72 UBEM studies, finding that EnergyPlus is the most widely used simulation engine (38\%) and CityGML the most common data model (27\%), though direct compatibility between the two remains limited. 
They observed that only 44\% of studies validated against measured data and that roughly 95\% lacked reproducibility due to insufficient data transparency \citep{MALHOTRA2022108552}. 

Hong \textit{et al.} (2016) developed the City Building Energy Saver (CityBES), an open web-based platform that automatically generates and simulates urban building energy models based on GIS data, integrating EnergyPlus simulations for city-scale retrofit analysis. 
Their approach significantly reduced technical barriers for municipalities by providing an accessible tool for evaluating energy conservation measures across entire building stocks \citep{hong_2016}.

Similarly, the City Energy Analyst (CEA) provides a computational framework for the analysis and optimization of energy systems in neighborhoods and city districts, integrating time-dependent simulations of demand, local resources, and distributed generation within a single spatial interface. 
This holistic approach underscores the importance of coupling urban design with energy system optimization to achieve sustainable development goals \citep{FONSECA_2016}.

Kasmeridis \textit{et al.} (2025) validated the BIPV-city platform, showing strong agreement with TRNSYS simulations (irradiance errors within $\pm2\%$, PV output deviations $<\pm10\%$), confirming that open digital twin tools can accelerate carbon-neutral urban transitions \citep{urbansci_kasmeridis_2025}.

Simulation results have also demonstrated the substantial impact of green infrastructure strategies on building energy performance.
For instance, Turhan \textit{et al.} (2025) illustrated that applying \textit{Hedera canariensis gomera} on walls and \textit{Phyllanthus bourgeoisii} on roofs reduced total energy use by 9.21\% and improved comfort by 23.21\%, underscoring the potential of nature-based retrofitting strategies \citep{urbansci7030096}.
Another illustrative example can be found in the work of Romeo \textit{et al.} (2025), which showed that combining envelope upgrades with photovoltaic systems could reduce annual energy consumption by up to 82\% and achieve near net-zero performance in Perth's residential sector, demonstrating the power of integrated passive and active measures \citep{urbansci9100421}.

In their analysis Rodrigues \textit{et al.} (2024) found that UBEMs effectively evaluate PV system performance and microclimatic impacts, finding that urban heat island effects can reduce PV efficiency and that BIPV configuration strongly affects cooling loads and outdoor comfort \citep{urbansci8040215}.

Several studies have investigated the importance of integrating UBEM with other urban systems, such as mobility or climate models, to address the complexity of urban processes and improve the robustness of models \citep{JOHARI2020109902, ABBASABADI2019106270}.
Khan \textit{et al.} (2025) demonstrated that integrating smart grids and renewable energy systems can cut urban energy use by up to 15\% and per capita emissions by 12\%, reinforcing that machine-learning-based prediction and digital infrastructure are vital for decarbonization \citep{urbansci_khan_2025}.

Despite the rapid progress that has been made and the important outcomes obtained, challenges remain, such as the need for computational efficiency when working with large urban datasets, the limited availability of high-resolution input, and uncertainty in occupant behavior modeling.
On this topic, Goy \textit{et al.} (2020) conducted a large-scale sensitivity analysis emphasizing that data quality and accessibility directly impact model accuracy and replicability \citep{en13164244}.

Johari \textit{et al.} (2020) observed that current UBEMs struggle to capture variations in building physics and occupant behavior, producing uncertainty in energy predictions. 
They found that hybrid approaches (i.e., combining physics-based and data-driven models) show the greatest potential, though validation against measured data remains limited \citep{JOHARI_2020}.

Chen \textit{et al.} (2020) addressed a similar issue by introducing a rapid automatic calibration method that learns from pre-simulated energy performance databases, showing that UBEMs can be effectively calibrated with limited annual energy data. 
Their approach reduced the average calibration effort to fewer than four simulation runs per building while substantially improving alignment with measured data \citep{CHEN2020115584}.

Building on the insights reported in the literature, this paper outlines the workflow implemented to conduct large-scale energy consumption simulations for individual buildings in the Municipality of Bologna, leveraging high-performance computing (HPC) resources.
In the following sections, we first provide a comprehensive overview of the datasets used as the foundation for running the energy simulations. 
We then describe in detail the methodology and computational tools employed to scale and orchestrate the simulation workflow, enabling the efficient execution of energy models for up to 25,000 distinct buildings. 
Finally, we present the results obtained from the large-scale simulations and discuss possible improvements and future directions for enhancing the scalability, accuracy, and applicability of the simulation framework.

This work is part of the project ``Bologna Digital Twin``, a collaborative initiative promoted by the Municipality of Bologna and coordinated by the Fondazione Bruno Kessler (FBK) in partnership with Cineca, the University of Bologna, and the Fondazione IU Rusconi Ghigi.

%%%%%%%%%%%%%%%%%%%%%%%%%%%%%%%%%%%%%%%%%%
\section{Building Data and Input Parameters}

The accuracy of a UBEM primarily depends on the quality and quantity of available data that are synthesized in the input file of the simulations, such as building geometries, construction materials, installed HVAC systems, occupancy types, and historical energy consumption data. When implementing a city-scale model, it becomes necessary to find a balance between the number of buildings modeled and the level of detail of the available data. 
Our prototype energy model has limited detail on individual buildings because there is no consistent, in-depth data available on the entire urban building stock. We only used available open data and constructed a system of building archetypes based on year of construction. This approach allowed us to simulate a very large number of buildings and provide a complete representation of urban consumption and its distribution across the territory. 

There are several tools for energy building simulations; among the various alternatives, the choice is to use EnergyPlus \citep{EnergyPlus940}, one of the most widely used simulation engines in major energy simulation tools such as CityBES \citep{hong_2016} and UMI \citep{UMI}.
EnergyPlus requires input files (\texttt{.idf} files) that specify the characteristics of each building.  At the end of the simulation, output files describing the “energy behavior” of the building are produced. 
Geometric information is extracted from open data provided by the Municipality of Bologna, and Digital Terrain Model (DTM) and Digital Surface Model (DSM) extracted from LiDAR data. Information on construction materials is obtained from the TABULA project and legal regulations.
The following paragraphs present the data used to prepare the input files for the simulations.

\begin{comment}
\subsection{Light Detection and Ranging
	Airborne (LiDAR)}
LiDAR is a remote sensing technology that uses laser pulses emitted from aircraft to measure the Earth’s surface and generate high-resolution three-dimensional representations of terrain and built environments. In our study, we make use of airborne LiDAR data collected over the city of Bologna, which has been processed and organized into 647 tiles, each covering a $500 m^2$ area. On average, each tile contains approximately 7 million points, corresponding to a point density of about 30 points per square meter, allowing for detailed spatial analysis of urban morphology.

In addition to the raw point clouds, we also use derived geospatial products such as the Digital Surface Model (DSM) and Digital Terrain Model (DTM), both provided at a $0.5 m$ spatial resolution. The DSM represents the elevations of natural and built features (vegetation, buildings), while the DTM captures the underlying bare-earth surface. These datasets enable a wide range of urban analyses, including building height estimation, shadowing, and energy modeling.

The LiDAR campaign is conducted annually, and for Bologna, we currently have data available for the years 2022, 2023, and 2024, allowing for temporal comparisons and the monitoring of urban evolution over time.

\begin{figure}[H]
	\centering
	%\includegraphics[width=\linewidth]{img/dsm_dtm.png}
	\caption{ Representation of a DSM (above) and DTM (below). \label{fig1}}
\end{figure}
\unskip
\end{comment}
\subsection{Open Data Comune di Bologna}
\begin{figure}[H]
	%\isPreprints{\centering}{} % Only used for preprints
	\centering
	\includegraphics[height=6.0 cm]{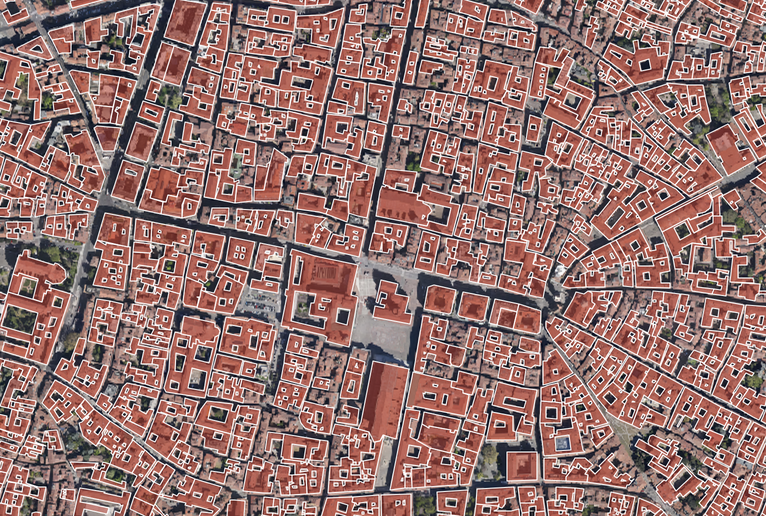}
	\caption{A set of shapefile polygons representing building footprints in Bologna from Edifici Particellari dataset.\label{fig1}}
\end{figure}

The \textit{Open Data Comune di Bologna} portal \footnote{Available at \url{https://opendata.comune.bologna.it/pages/home/}.}, managed by the Municipality of Bologna, provides a comprehensive repository of datasets related to different aspects of the city. Among the available datasets, we utilize the following key resources:
\begin{itemize}
	\item \textit{Edifici Particellari}: This dataset offers detailed information on individual building parcels within Bologna, including attributes such as plan area and Geo Shape of buildings. Figure ~\ref{fig1} shows a visualization of the buildings' geometries extracted from this dataset.
	\item \textit{Edifici Volumetrici}: This dataset provides three-dimensional representations of buildings, capturing volumetric data such as height, plan area and Geo Shapes of buildings.
	\item \textit{Numeri Civici}: This dataset contains the geolocated addresses (house numbers) throughout the city, facilitating accurate mapping and navigation tasks, and a variable that has been used as proxy for construction year.
\end{itemize}

\unskip

\subsection{TABULA}
In our study, we also integrate information from TABULA (Typology Approach for Building Stock Energy Assessment) \citep{tabula}, an European research project aimed at creating a standardized typology of residential buildings across EU countries. 
The project provides reference data for building components, including typical wall constructions, insulation levels, and thermal transmittance values (U-values) based on building age, typology, and country-specific construction practices.

We use the Italian TABULA dataset \footnote{Available at \url{https://webtool.building-typology.eu/}.} to estimate parameters such as wall thickness, thermal conductivity, and overall heat transfer coefficients for the building stock in Bologna.
This information is essential for urban energy simulations, particularly when detailed architectural or construction data are not available for each individual building. 
By leveraging TABULA, we ensure a consistent and replicable approach to modeling the thermal performance of buildings at scale across the urban fabric.

\subsection{Legal Regulations}
In addition to TABULA, we also refer to Italian national building codes and energy efficiency regulations to verify thermal parameters. These normative references are crucial for aligning our assumptions with the current legal framework, especially for newly built or renovated structures where regulation-based requirements may differ from historical averages provided by TABULA.

By combining empirical typologies with regulatory standards, we aim to ensure both accuracy and legal consistency in our energy modeling approach. This is a list of used regulations:

\begin{itemize}
	\item UNI 10351 - Valori di trasmittanza, conducibilità termica e vapore di materiali OMOGENEI;
	\item UNI EN ISO 6946 ITA - Resistenza e trasmittanza termica di elementi dell’edilizia.
\end{itemize}

\subsection{Energy Plus and Climate File}
EnergyPlus is an open-source building energy simulation software that has been produced by the U.S. Department of Energy since 1997. It models heat transfer through the building envelope, the operation of HVAC systems, and natural or mechanical ventilation, while also accounting for solar gains, infiltration, and occupant behavior. 
To perform these simulations, EnergyPlus requires two essential inputs: (1) a building description file (i.e., the \texttt{.idf} file) containing information such as geometry, materials, internal loads, and system configurations, and (2) a climate file (\texttt{.epw}) that provides meteorological data representative of the study area.

The \texttt{.epw} file (EnergyPlus Weather file) is a standardized format that contains hourly data on temperature, humidity, solar radiation, and other climatic variables. By using Typical Meteorological Year (TMY) datasets EnergyPlus is able to simulate energy performance under realistic local conditions. 
Our prototype incorporates the TMY 2009-2023 for Bologna \footnote{Available at \url{https://climate.onebuilding.org/WMO_Region_6_Europe/ITA_Italy/index.html}.}. This allows the model not only to estimate annual consumption and emissions but also to test “what-if” scenarios, compare retrofit strategies, and evaluate building performance under different climate assumptions. 
In this way, the climate file plays a crucial role in ensuring that simulations are both location-specific and reproducible.

\begin{comment}
	\footnote{Available at \url{https://webtool.building-typology.eu/#bm}}
\begin{figure}[H]
	%\isPreprints{\centering}{} % Only used for preprints
	\includegraphics[width=15.0 cm]{img/file_climatico.png}
	\caption{Average Hourly Temperature in Bologna (TMY 2009–2023) of .epw climate file used for EnergyPlus simulations.\label{fig2}}
\end{figure}
\unskip
\end{comment}

%%%%%%%%%%%%%%%%%%%%%%%%%%%%%%%%%%%%%%%%%%
\section{Methodology}
\begin{figure}[H]
	%\isPreprints{\centering}{} % Only used for preprints
	\centering
	\includegraphics[height=6.0 cm]{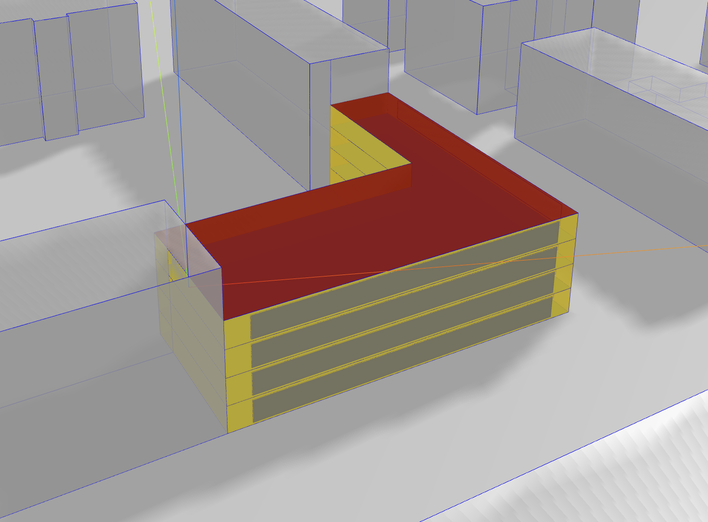}
	\caption{Example of $2.5D$, or LoD1, geometry of a building in Bologna using building plan vertices taken from its Geo Shape and the height found by LiDAR. The windows were added respecting the legal regulation between $\frac{windowed_{area}}{plan_{area}} = \frac{1}{8}$.\label{fig2}}
\end{figure}

Our strategy involves collecting and integrating geospatial and typological data from multiple sources to characterize the geometrical and thermal properties of buildings in the urban context.
Some properties, such as height and plan area, are assigned directly to individual buildings, while others, like building material and window characteristics, are specified after partitioning buildings into archetypes based on TABULA values.
Following this step, we implemented a pipeline to automatically generate a specific \texttt{.idf} file containing all the information for each building.
The \texttt{.idf} files are then used as input for simulations carried out with EnergyPlus. 
Figure \ref{fig2} shows the visualization of an input file that was used for the simulation.

Leveraging the Leonardo high-performance computing (HPC) infrastructure it was possible to perform the simulation of approximately 25,000 buildings in less than 30 minutes using 1,120 CPU cores.
After the simulation of the current buildings state, we analyzed possible retrofitting scenarios using optimization techniques such as Pareto analysis, and identified the archetypes with the higher benefits concerning energy consumption.

The following sections describe in depth how the homogeneous dataset for the building stock was created.

\subsection{Data Integration}
Geometrical data was primarily sourced from \textit{Open Data Comune di Bologna} portal, which provides detailed footprints and approximate building heights. LiDAR data were used to fill gaps in the Open Data, providing accurate height information. 
 
The dataset \textit{Edifici Particellari} and the dataset \textit{Edifici Volumetrici} have different identification codes (as they refer to different objects: within the same parcel there may be one or more volumetric units); for this reason, it was necessary to perform a spatial join in order to integrate them. 
The \textit{Numeri Civici} dataset, on the other hand, shares the same identification code as the \textit{Edifici particellari}, although it has significantly fewer unique codes than the latter.  
The integration of these different sources made it possible to build a comprehensive dataset of approximately 25,000 building constructions, using the parcel building code as the identifier. The information contained in this dataset for each unit includes geometry, surface area, height, and year of construction. 

\unskip

\subsection{Building Height Extraction}
\begin{figure}[H]
	%\isPreprints{}{% This command is only used for ``preprints''.
	%\begin{adjustwidth}{-\extralength}{0cm}
		\centering
		%} % If the paper is ``preprints'', please uncomment this parenthesis.
		\subfloat[]{\includegraphics[height=7.0cm]{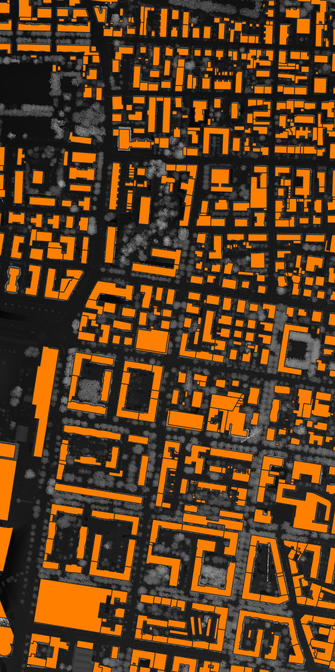}}
		%\hfill
		\subfloat[]{\includegraphics[height=7.0cm]{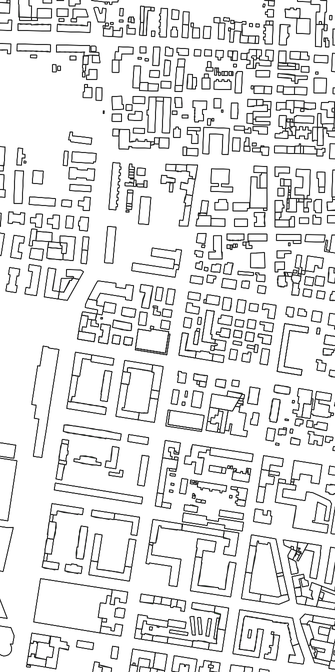}}\\
		%\isPreprints{}{% This command is only used for ``preprints''.
	%\end{adjustwidth}
	%} % If the paper is ``preprints'', please uncomment this parenthesis.
	\caption{Example of buildings height extraction. a) Building shapes from Open Data. b) Building perimeter footprint used to extract the height. \label{fig3}}
\end{figure}

When data from the \textit{Edifici Volumetrici} dataset was not available, building heights were extracted by using polygon shapefiles representing building footprints obtained from \textit{Open Data Comune di Bologna} portal. 
To accurately determine building heights, we calculated the difference between Digital Surface Model (DSM) and Digital Terrain Model (DTM), thus excluding ground elevation. Given that EnergyPlus considers buildings as 2.5D structures (box-shaped without roofs, so LoD1 buildings), only the perimeter of each building polygon was considered. Polygons were buffered to ensure the exclusion of roof heights, and the maximum DSM-DTM difference along each buffered perimeter was selected to represent the building’s height.
Figure \ref{fig3} shows an example of buildings height extraction.

\subsection{Building Archetypes}
\begin{figure}[H]
	%\isPreprints{\centering}{} % Only used for preprints
	\centering
	\includegraphics[height=6.0cm, trim=50 200 0 50, clip]{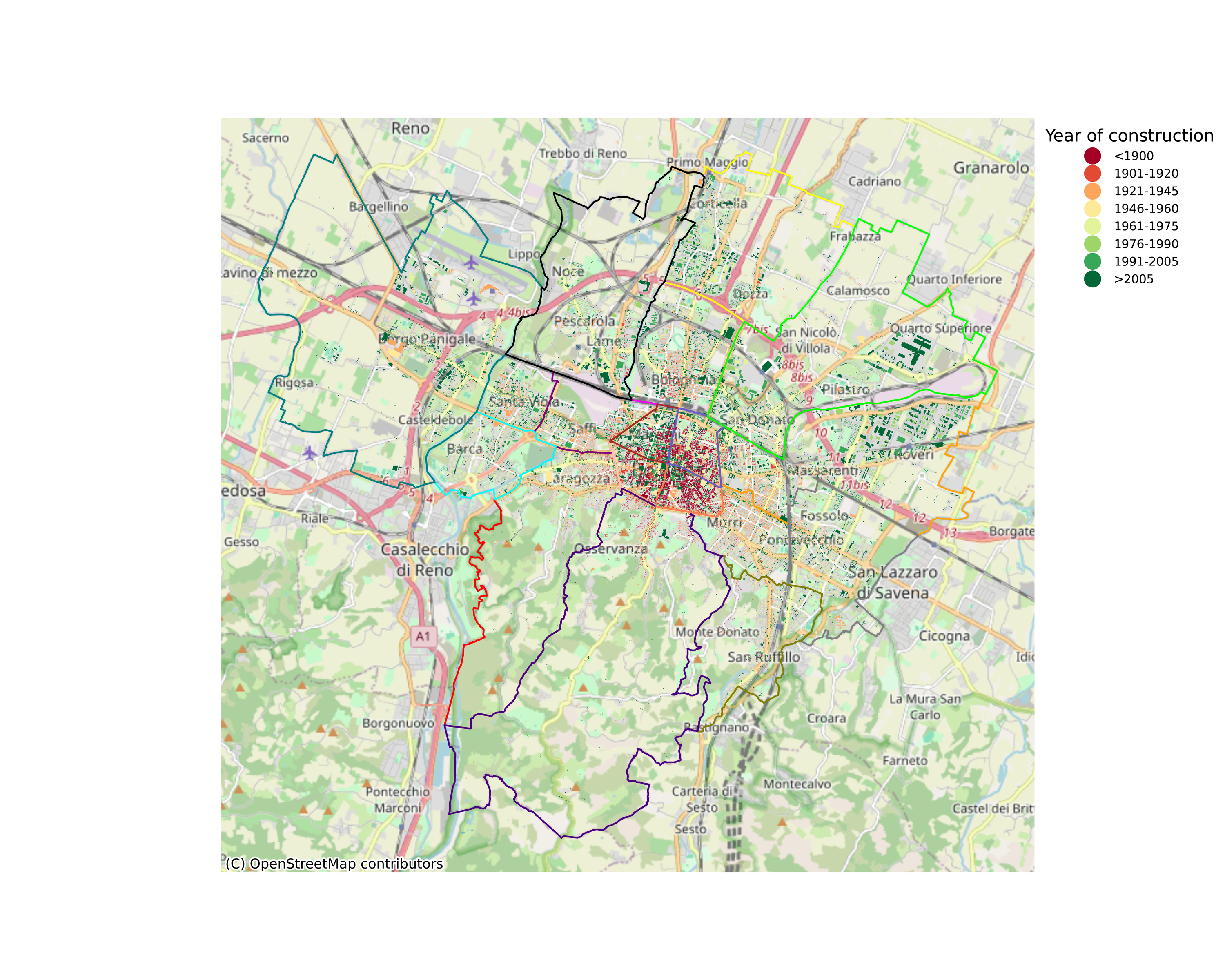}
	\caption{Age-based archetypes distribution in the city center of Bologna.\label{fig4}}
\end{figure}

In order to further differentiate buildings, we grouped buildings into archetypes according to a range of construction years referencing the TABULA classification. Eight types of archetypes were identified based on the year of construction (<1900, 1901-1920, 1921-1945, 1946-1960, 1961-1975, 1976-1990, 1991-2005, >2005) distributed across the areas of Bologna as illustrated in Figure \ref{fig4}. For each category, we enrich the corresponding input files with information provided on building components, including wall construction, insulation levels, thermal transmittance values (U-values), window composition, etc.,

\unskip

\section{Results}

The simulation pipeline comprises two sequential stages: the generation of the EnergyPlus input file (\texttt{.idf}) and the execution of the EnergyPlus simulation itself.
Managing the parallel execution of this pipeline across tens of thousands of buildings poses substantial orchestration challenges as simulations must be dynamically queued and dispatched as computational resources and ``logical``\footnote{A simulation cannot be started before the corresponding \texttt{.idf} file is created.} dependencies become available, ensuring continuous utilization of CPU cores without manual intervention.

Each building simulation is largely independent from other construction simulations, with the only interdependence arising from shadow interactions between buildings.
These interactions are handled by embedding the geometry of neighboring buildings directly within the input file of the target building.
Beyond this preprocessing step, simulations operate in complete isolation, without requiring inter-process communication or synchronization.
This independence classifies the task as an embarrassingly parallel problem—one that is particularly well suited for high-performance computing environments.
Once the \texttt{.idf} files are generated, the workload can be readily distributed across multiple compute nodes, significantly reducing total simulation time.

To address these orchestration and scalability challenges, we adopted the Ray distributed computing framework, implementing each step as an independent Ray task.
This architecture enables fault-tolerant and resource-efficient orchestration while providing excellent scalability: by simply adding compute nodes to the Ray cluster, the system can seamlessly execute a larger number of simulations in parallel, thereby reducing total computation time in proportion to the available resources.

For optimal resource utilization, we assigned one CPU core per task.
This decision was guided by the current design of EnergyPlus, which does not support multithreading within a single simulation.
As a result, allocating multiple cores to a single simulation yields no performance gains.
Consequently, distributing the workload such that each simulation runs on a single core maximizes the resource usage.

The EnergyPlus simulations were executed on the Leonardo high-performance computing (HPC) cluster at Cineca.
Each compute node provides 112 CPU cores and 512 GB of RAM, offering a solid foundation for single-node parallelization of the simulation workload.
By leveraging 10 compute nodes—equivalent to 1,120 cores—the simulation of approximately 25,000 buildings was completed in roughly 30 minutes.
For further discussion of scalability aspects, see Section \ref{simul_performance}.

\begin{figure}[H]
	%\isPreprints{\centering}{} % Only used for preprints
	\centering
	\includegraphics[height=6.0cm]{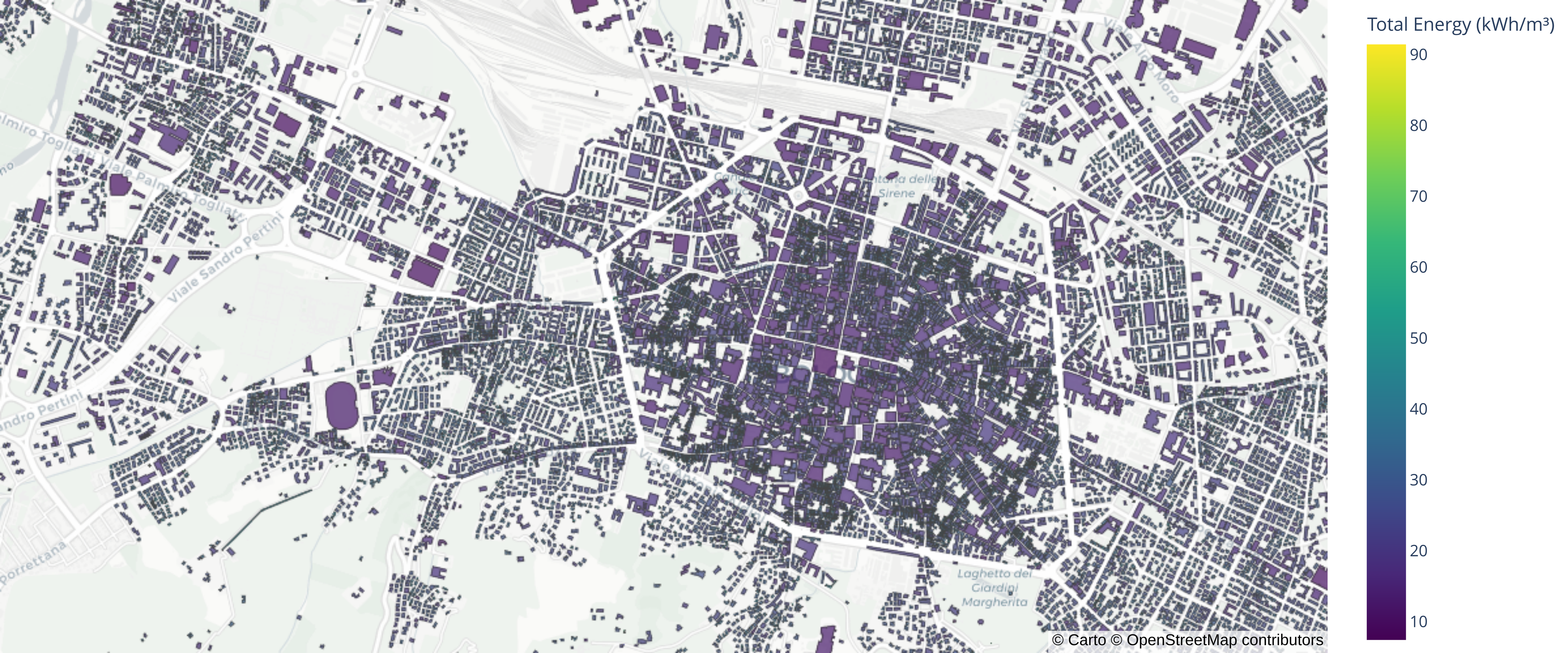}
	\caption{Annual normalized simulated energy consumption for buildings in Bologna center. \label{fig6}}
\end{figure}

Figure \ref{fig6} shows the result of the simulation step.

\unskip

\subsection{Calibration}
To evaluate the reliability of the archetypes and the predictive accuracy of the resulting simulations, a calibration phase must be conducted using real energy consumption data from the residential buildings.
However, consumption data with daily or monthly resolution are generally difficult to access due to privacy constraints and regulatory restrictions, making their use challenging unless the data are properly anonymized and aggregated.
Moreover, such datasets are often held by multiple private entities, further complicating efforts to collect the comprehensive information required for robust calibration.

To compensate the lack of direct observational data on energy consumption in Bologna, we verified the reliability of our simulations by comparing the results with the energy demand of archetypal buildings provided by the TABULA project for the city of Turin.

However, two important considerations must be noted regarding this comparison:

\begin{itemize}
	\item The values provided by TABULA are based on the climate of Turin.
	      Although both Bologna and Turin belong to the same Italian climate zone (Zone E), Bologna generally experiences milder temperatures, which should result in lower heating energy consumption.
	      A potential approach to address this discrepancy is to run the simulations using a weather file representative of Turin, thereby estimating the average heating demand per square meter under the same climatic conditions, enabling a consistent comparison with the TABULA data;

	\item According to \citep{tabula} the heating needs reported in the TABULA web tool are computed based on the UNI/TS 11300 series of technical specifications, which adopt a \textit{quasi-steady-state} method.
	      This method approximates energy flows as constant over time.
	      In contrast, Energy Plus uses a dynamic method that simulates energy consumption by accounting for continuous changes that influence energy use.
		  A few studies \citep{CORRADO2016200,CORRADO2018} show that UNI/TS 11300 procedure yields higher estimated energy needs than those simulated with EnergyPlus.
	      Consequently, it is reasonable to expect EnergyPlus to yield lower heating energy consumption estimates than those reported by TABULA.
\end{itemize}

\begin{figure}[H]
	%\isPreprints{\centering}{} % Only used for preprints
	\centering
	\includegraphics[width=0.8\textwidth, keepaspectratio]{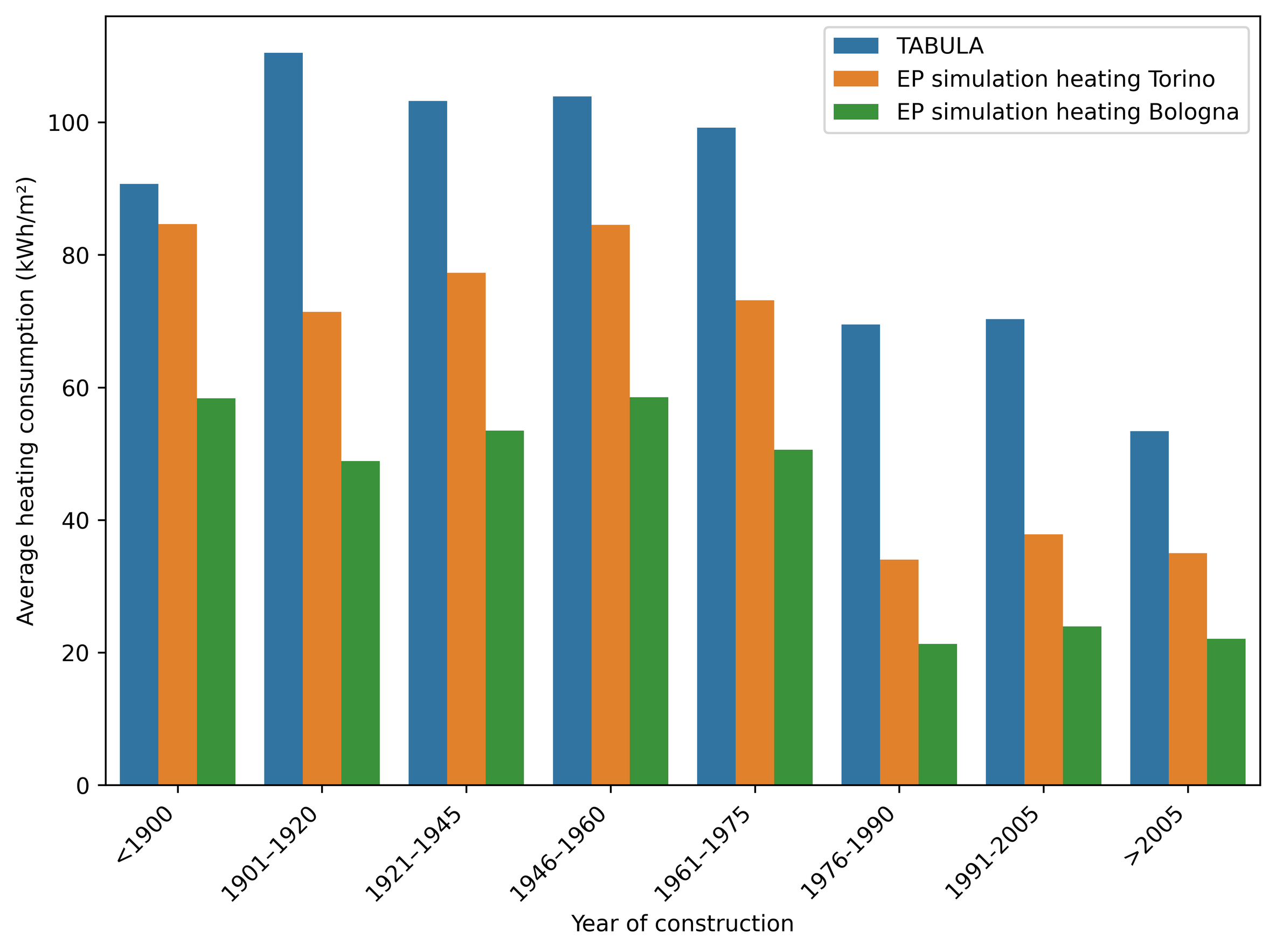}
	\caption{Comparison between the heating consumption simulated by Energy Plus for Bologna, Torino and the energy needs reported by TABULA.\label{figB.1.1}}
\end{figure}

The plot in Figure~\ref{figB.1.1} compares the results obtained for Bologna and Turin using EnergyPlus and TABULA values.

\unskip

To calculate the average energy consumption per square meter for the EnergyPlus simulation, we averaged the heating and cooling requirements of buildings with similar floor areas to those used to construct the archetypes by Tabula.

Consistent with the findings reported in previous studies, the heating consumption estimated using EnergyPlus is lower than the heating demand calculated according to UNI/TS 11300 in TABULA.
Nevertheless, the overall trend remains comparable, indicating that the simulation results are reasonably reliable.
For a detailed description of the simulation results, see Appendix \ref{current_energy_consumption}.

\subsection{Scenario Analysis}
In contexts where a proper calibration is missing, simulations can still be leveraged for qualitative ``what-if`` scenario analyses.
These analyses explore hypothetical interventions and their potential energy impact across the building stock.

Typical scenarios analyzed include:
\begin{itemize}
	\item Neighborhoods benefit: which neighborhoods would benefit most from efficiency policies.
	\item No renovation vs. full retrofit: assessing the citywide performance improvement based on building's envelope improvement and window replacement;
\end{itemize}

While qualitative, these scenarios provide valuable foresight for long-term urban planning, particularly in districts under redevelopment pressure or policy-driven transformation goals.

\subsubsection{Impact of Building Renovations on Neighbourhood Energy Performance}
\begin{figure}[H]
	%\isPreprints{\centering}{} % Only used for preprints
	\centering
	\includegraphics[width=0.8\textwidth, keepaspectratio]{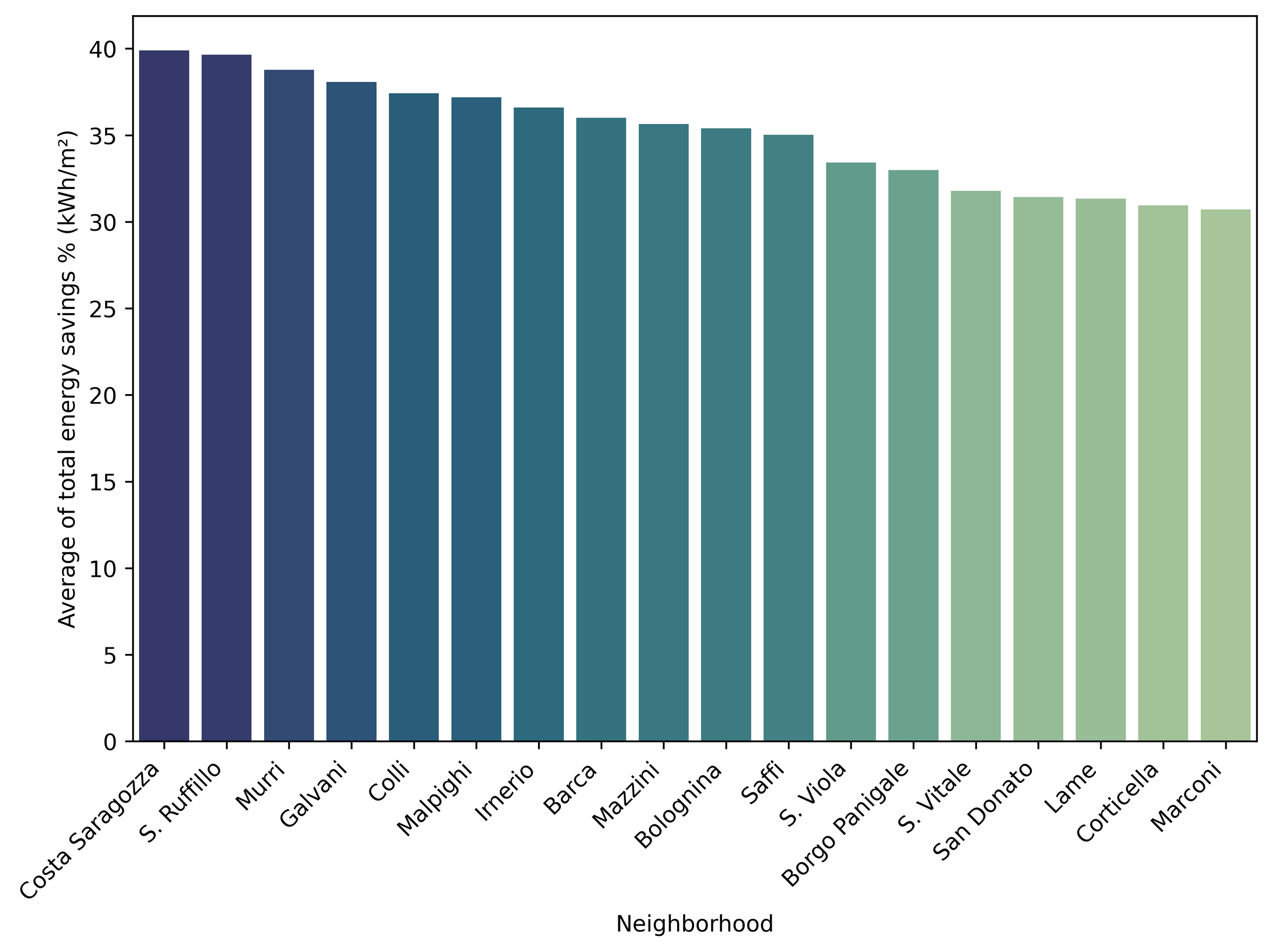}
	\caption{Total energy savings percentage (in $kWh/m^2$) grouped by neighborhoods.}
	\label{energy_savings_by_neighborhood}
\end{figure}

For each archetype, TABULA proposes two types of refurbishment. We choose the standard one and applied it to all the archetypes; then, for each neighborhood, we calculated the average total energy savings percentage in $kWh/m^2$ achievable across the buildings in that area.
As it can be seen in Figure \ref{energy_savings_by_neighborhood} there isn't a restricted subset of neighborhoods that would benefit the most in terms of energy savings.
This because old buildings are evenly spread across all the neighborhoods.

\unskip

\subsubsection{City-Scale Effects of Building Retrofits}
\begin{figure}[H]
	%\isPreprints{\centering}{} % Only used for preprints
	\centering
	\includegraphics[width=0.8\textwidth, keepaspectratio]{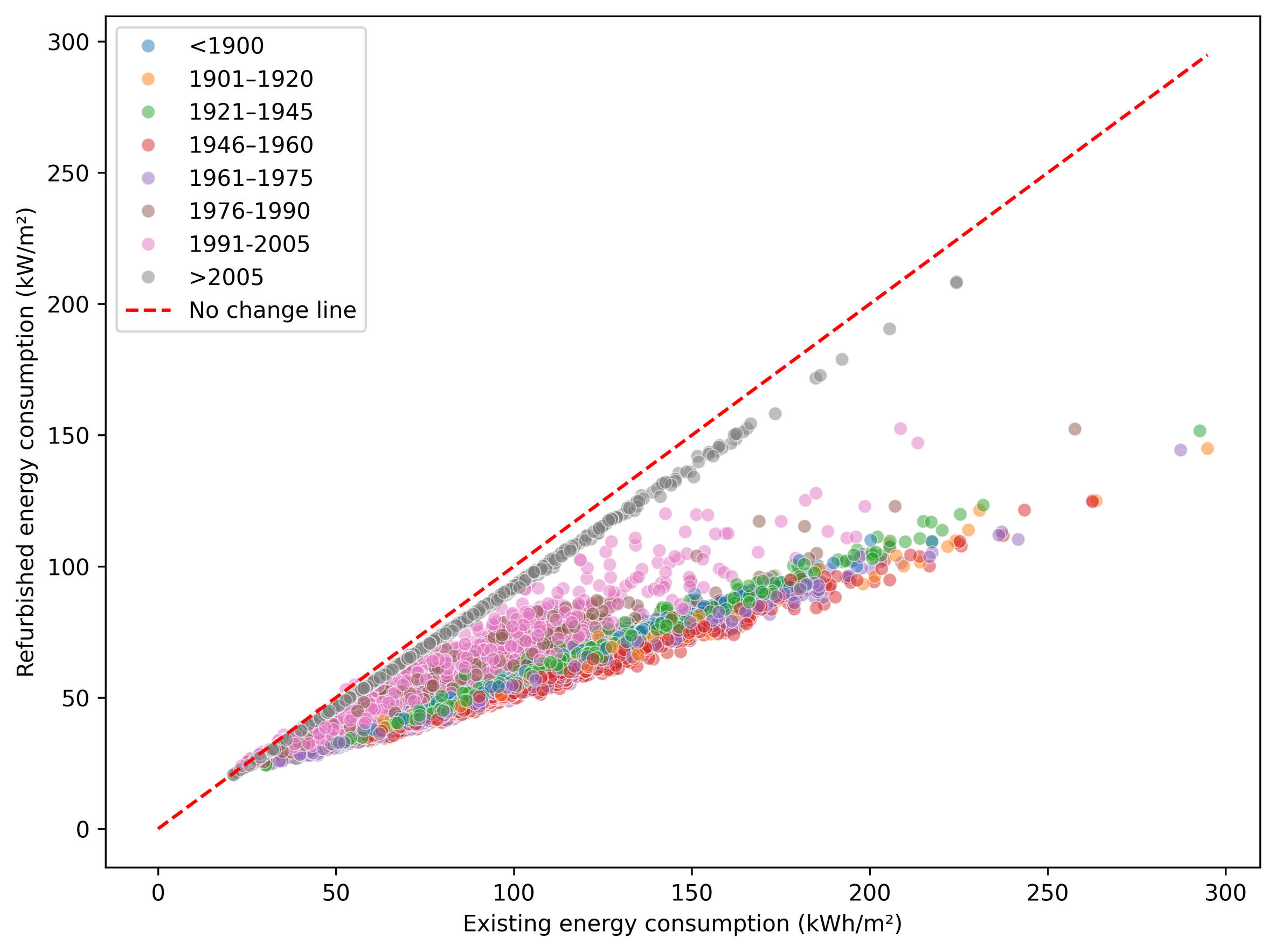}
	\caption{Assessment of the refurbishment impact grouped by archetype.\label{figD.1.3}}
\end{figure}

Figure~\ref{figD.1.3} shows the energy consumption of buildings before and after the refurbishment.
Each point represents a building. The x-axis shows the energy consumption of the existing state, and the y-axis shows the energy consumption after standard refurbishment.
The closer a building is to the red line, the less impact the refurbishment will have on it.
As it can be seen, the most recent buildings are very close to the red line because the construction materials used are already efficient enough.
On the opposite, archetypes from the middle of the past century offer the best advantages in terms of consumption improvement.

\unskip

Refurbishment planning often involves trade-offs between multiple conflicting objectives, such as: reducing energy consumption, minimizing investment costs, and limiting the number of buildings targeted for intervention.

In such multi-objective contexts, the Pareto front is a valuable decision-making tool.
A Pareto front represents the set of non-dominated solutions (i.e., scenarios in which no objective can be improved without worsening another); in other words, rather than focusing on a single optimal solution, the front highlights a range of optimal trade-offs.
In the context of refurbishment, this might mean identifying scenarios where maximum energy savings are achieved for a given cost, or where the highest return is obtained from refurbishing the fewest buildings.

For our purpose, the optimization problem consists of minimizing the following objectives:
\begin{itemize}
	\item number of buildings refurbished;
	\item total energy consumption of the city.
\end{itemize}

The explored scenarios are based on combinations of refurbished buildings based on the construction periods.
Specifically, in each scenario all buildings from certain periods are retrofitted (i.e., 0\% or 100\% of buildings made in 1950).

\begin{figure}[H]
	%\isPreprints{\centering}{} % Only used for preprints
	\centering
	\includegraphics[width=0.8\textwidth, keepaspectratio]{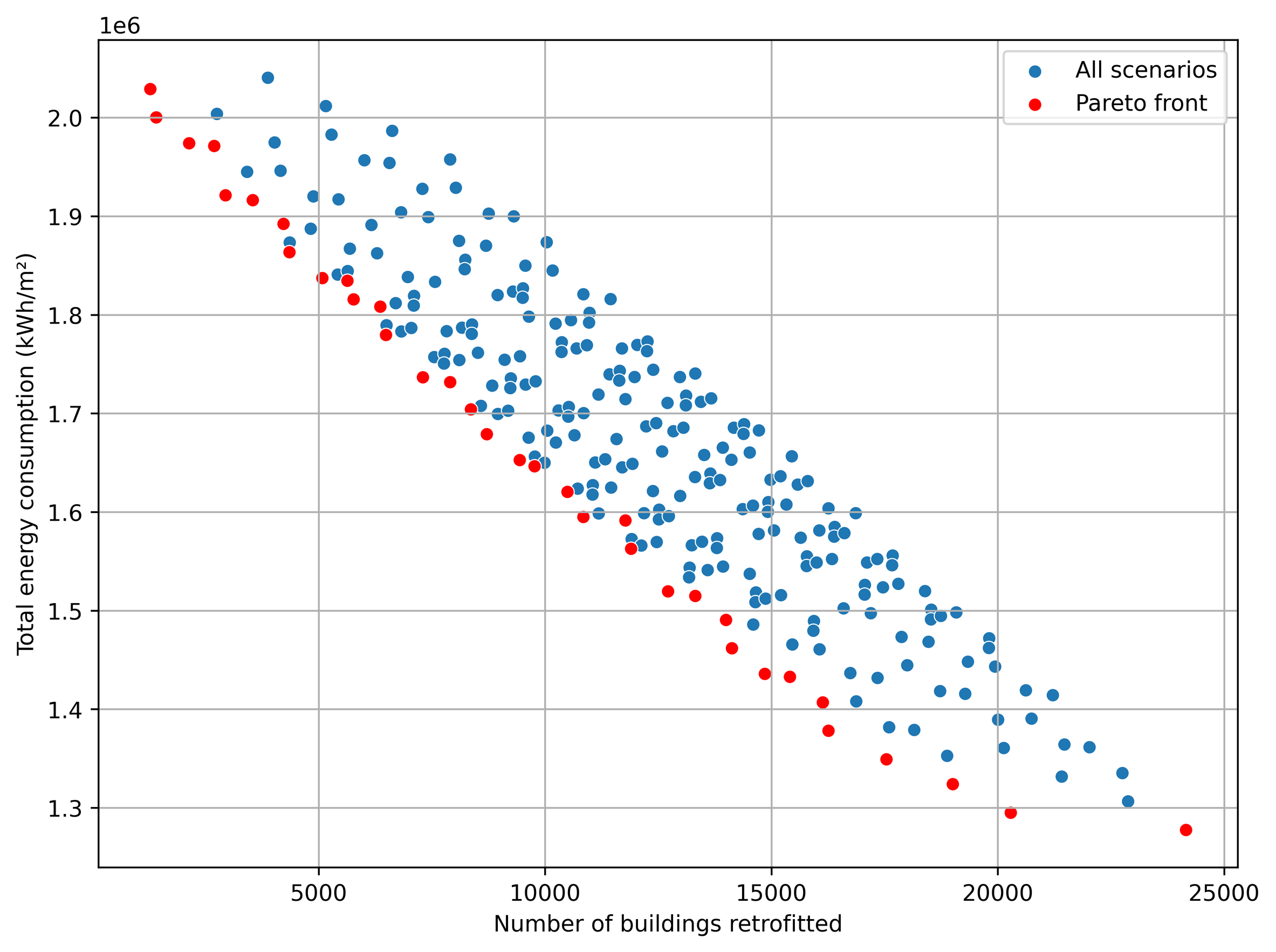}
	\caption{Pareto front for refurbishment scenario selection.\label{figD.1.4}}
\end{figure}

Figure~\ref{figD.1.4} shows the results of the refurbishment scenario selection.
Each point represents a specific scenario: the red points describe optimal scenarios in terms of minimal number of buildings retrofitted and the total energy per $m^2$ used; the light blue bue points represent non-optimal scenarios.

\unskip

\begin{figure}[H]
	%\isPreprints{\centering}{} % Only used for preprints
	\centering
	\includegraphics[height=5.0cm]{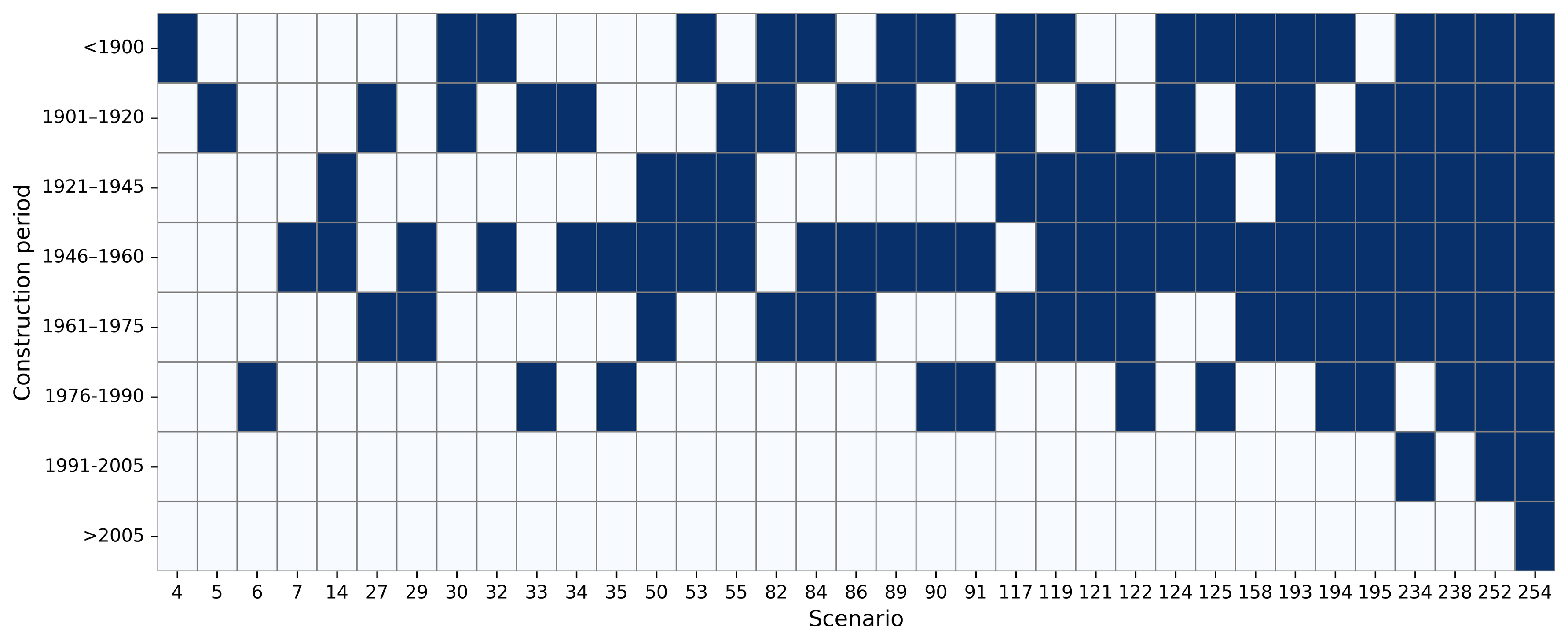}
	\caption{A binary visualization of the retrofit combinations for scenarios on the Pareto front.\label{figD.1.5}}
\end{figure}

In Figure~\ref{figD.1.5}, each column of the binary map  represents a scenario on the Pareto front.  A scenario is defined as a combination of archetypes that have been retrofitted and archetypes that have been left unchanged. There is therefore a one-to-one correspondence between the red dots in Figure \ref{figD.1.4} and the columns of the matrix in Figure \ref{figD.1.5}.
Each row refers to a specific archetype: a blue cell indicates that the archetype in the corresponding row has been completely retrofitted in the scenario in the corresponding column. A white cell indicates that the corresponding archetype has not been modified in that scenario.
This helps visually detect patterns, for example: “Buildings from 1946 to 1960 are almost always retrofitted in solutions that turned out to be optimal” or “>2005 buildings are never retrofitted in optimal scenarios”.

\unskip

As evidenced from the binary map, the archetypes that are often included in the Pareto front are: 1946-1960 (27 times), 1901-1920 (20 times), <1900 (19 times), 1961-1975 (18 times).
Buildings corresponding to these archetypes thus constitute the key priority areas for decision-makers.

\section{Discussion}
The use of geometric data, complemented with LiDAR data and TABULA archetypes, enabled the development of city-scale simulations starting from publicly available datasets.
The HPC infrastructure proved effective in handling the computational load associated with large scale simulations, providing a robust and scalable scalable solution that can be replicated in other urban contexts.

However, several limitations must be acknowledged.
To enable simulation at the urban scale, only a limited set of non-standard, building-specific parameters could be used, resulting in simplifications regarding construction characteristics and operational behavior (e.g., HVAC system configurations, occupancy patterns, etc.).
These simplifications, while necessary for a simulation at a large scale, can introduce uncertainties in the results, particularly when assessing buildings with atypical characteristics or complex operational profiles.
One of the most critical and uncertain aspects is the behavior of building occupants.
Occupant schedules, internal gains, and system usage patterns have a significant impact on energy consumption, but are inherently variable and difficult to capture accurately in large-scale archetype-based models.

Although archetypes provide a structured and standardized foundation for assigning construction materials, they cannot fully reflect the diversity of real-world conditions at the individual building level.
This limitation becomes especially evident when key information, such as the construction year or records of past refurbishments, is unavailable or only partially available, as in our case.

Finally, the absence of a proper calibration step limits the reliability of the results, potentially leading to discrepancies between simulated outcomes and actual energy performance.
As a consequence, the findings should be interpreted primarily in qualitative terms rather than as precise quantitative estimates.

\section{Conclusions}
This study successfully demonstrates the feasibility of large-scale Urban Building Energy Modeling (UBEM) by combining publicly available datasets, regulatory data, real world archetypes from TABULA, and the computational power of a high-performance computing (HPC) environment.
Through the integration of detailed geospatial datasets—including LiDAR-derived DSM/DTM, building footprints, and volumetric data—with archetypal characteristics, we have constructed a robust and scalable pipeline capable of simulating the energy behavior of over approximately 25,000 buildings in the city of Bologna in less than 30 minutes.

The use of EnergyPlus enabled detailed thermal and energy simulations under realistic climate conditions, while the adoption of a parallel architecture based on Ray facilitated the efficient orchestration and execution of thousands of independent simulations.
The results provide an unprecedented, spatially understanding of energy consumption patterns in Bologna’s urban fabric and offer a valuable tool for supporting energy planning and decarbonization strategies.

Looking ahead, future improvements should focus on the integration of real-world consumption data, which should be anonymized and aggregated appropriately, to calibrate the model and enhance reliability of the simulations.
Introducing temporal dynamics, such as variable occupancy patterns and seasonal effects, would further improve the realism of the simulations.
Moreover, integrating this modeling framework with other urban layers, including mobility data, socio-economic indicators, and urban microclimate models, would facilitate the development of more comprehensive urban energy strategies.

In conclusion, this research demonstrates how a scalable data-driven UBEM framework can support the digital transformation of urban energy systems, providing a replicable methodology for cities aiming to transition toward sustainable and climate-resilient futures.

%% Optional
\newpage
%%\appendixtitles{yes} % Leave argument "no" if all appendix headings stay EMPTY (then no dot is printed after "Appendix A"). If the appendix sections contain a heading then change the argument to "yes".
%%\appendixstart
\section*{Appendix}
\appendix
\section{Simulation performance evaluation} \label{simul_performance}
\subsection{Exploring the impact of shading buffers and cluster size on total simulation time}
In dynamic energy simulations, the representation of shading interactions between constructions is crucial for reliably estimating a building's energy consumption, particularly for heating and cooling loads.
EnergyPlus allows for the detailed modeling of shading geometries.
For our purposes, we used the \texttt{Shading:Building:Detailed} object, which describes shading elements that are external to the building\footnote{For more information about the shading module, you can refer to the official documentation \url{https://bigladdersoftware.com/epx/docs/8-6/input-output-reference/group-thermal-zone-description-geometry.html\#shadingsitedetailed-shadingbuildingdetailed}.}.
Therefore, we modeled nearby buildings using their geoshapes and included them into the \texttt{.idf} file of the building to be simulated.

The choice of shading radius, which defines the spatial extent of nearby buildings included in the simulation, affects both the accuracy and complexity of energy simulation: a smaller radius may underestimate shading effects, while a larger one captures more accurate interactions, especially in dense urban areas.
However, increasing the radius also raises the number of surfaces involved in ray-tracing and shadow projections computations, substantially increasing simulation times.

\begin{figure}[H]
	%\isPreprints{\centering}{} % Only used for preprints
	\centering
	\includegraphics[width=0.9\textwidth, keepaspectratio]{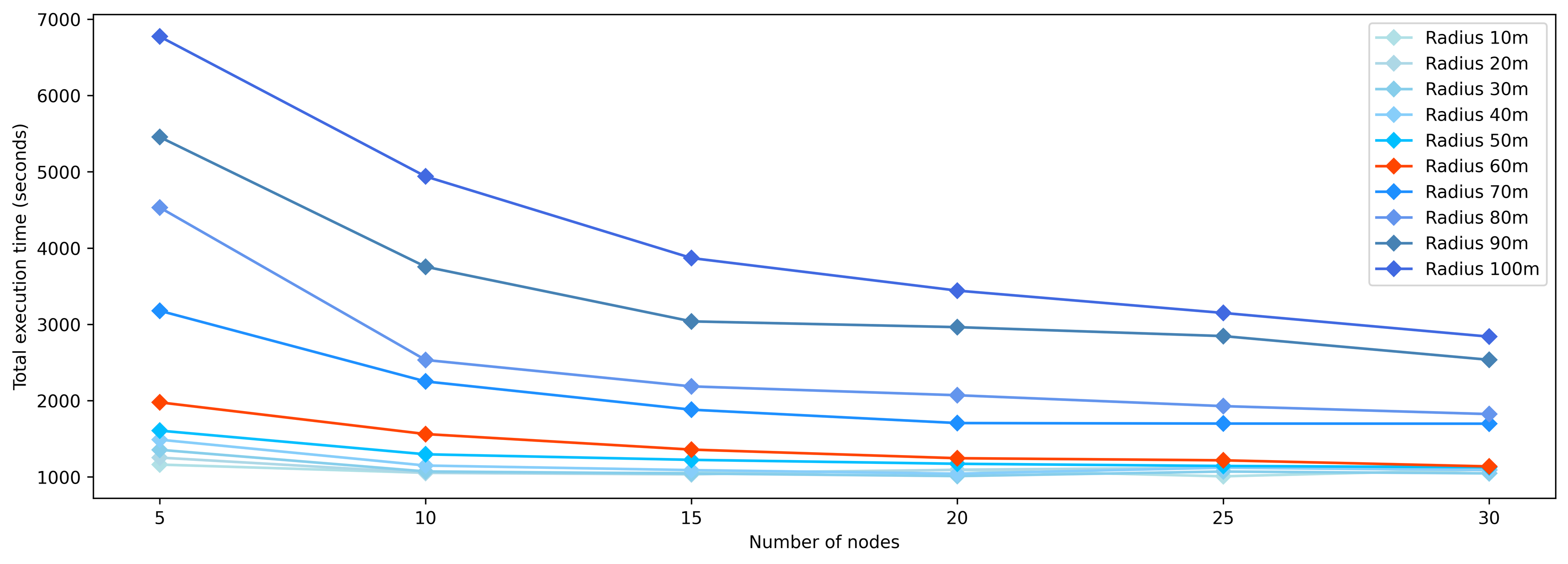}
	\caption{Total execution time variation reducing the radius and adding more nodes to the Ray
		cluster.\label{figA.1.1}}
\end{figure}

Our simulation pipeline therefore requires tuning of the shading radius parameter.
Figure~\ref{figA.1.1} presents the execution times associated with the simulation of the entire building stock while varying the number of nodes utilized for parallelization and adjusting the radius used to model surrounding buildings.
As evidenced by the plot, an increase in the inclusion radius leads to an increase in total simulation time.
This behavior manifests as a vertical translation along the time axis in curves corresponding to different radii.

\unskip

The total execution time associated with a given radius configuration can be reduced by increasing the number of simulations executed in parallel.
This is accomplished by augmenting the pool of Ray workers through the addition of computational nodes.
The Ray scheduler will then autonomously allocate the available CPU cores within the cluster to execute multiple simulations concurrently, thereby reducing the number of buildings remaining in queue waiting for resources.
However, it is important to note that even when utilizing a cluster with 24,537 CPU cores\footnote{The total number of buildings to be simulated is 24,537. Under this configuration, perfect parallelism is attained since a one-to-one correspondence exists between building simulations and available CPU cores.}, the overall execution time remains constrained by the duration of the longest individual building simulation.
Consequently, each radius configuration is associated with a baseline execution time that cannot be further reduced through parallelization.

\begin{figure}[H]
	%\isPreprints{\centering}{} % Only used for preprints
	\centering
	\includegraphics[width=0.8\textwidth, keepaspectratio]{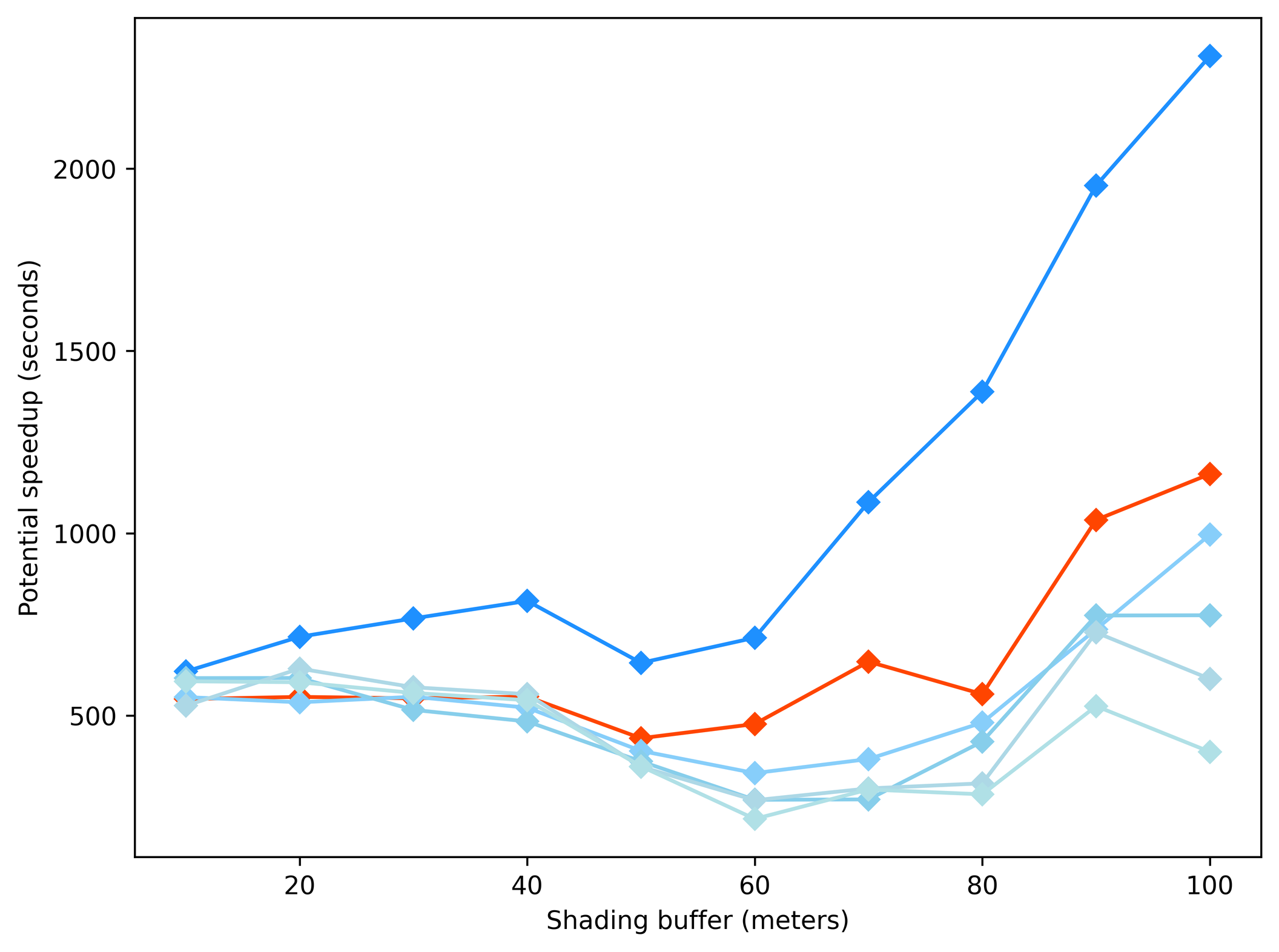}
	\caption{Potential speed-up obtainable by increasing the number of concurrent building
		simulations.\label{figA.1.2}}
\end{figure}

We evaluated the potential benefits of increased parallelization across different combinations of nodes and radii by subtracting the mean simulation time of the ten slowest-running buildings from the total simulation time.
This provides an estimate of the time savings achievable by executing a larger number of building simulations in parallel.
The resulting speed-ups obtained by progressively increasing the number of worker nodes in the computing cluster are illustrated in Figure~\ref{figA.1.2}.

\unskip

As shown in the Figure~\ref{figA.1.2}, increasing the number of nodes in the Ray cluster results in a downward shift along the vertical axis, as more buildings can be simulated concurrently, thereby reducing the overall execution time.
The spread between the curves is also impacted by variations in buffer size: larger buffers require a greater number of shade projections to be computed, which increases the baseline execution time and consequently diminishes the relative gains achievable through additional parallelization.

With a cluster configuration of 10 computing nodes, the performance improvement over the previous node configuration is maximized.
Although a further increase in the number of nodes could theoretically shorten the gap between the baseline and total execution times, the resulting gains are marginal when considering the additional resource consumption and the increase in scheduling and communication overhead.
Specifically, only negligible reductions in execution time are observed, especially for radii configurations smaller than 60m. Consequently, the configuration with 10 nodes and 60m buffer represents a balanced and robust option, achieving efficient utilization of computational resources while preserving sufficiently low total execution times.

\subsection[\appendixname~\thesubsection]{Sensitivity of energy demand to shading buffers}
The shadow buffer radius not only affects the simulation time, but also the total simulated energy consumption.
In this section, we investigate how varying the shading buffer radius affects simulated energy consumption.

\begin{figure}[H]
	%\isPreprints{\centering}{} % Only used for preprints
	\centering
	\includegraphics[width=0.8\textwidth, keepaspectratio]{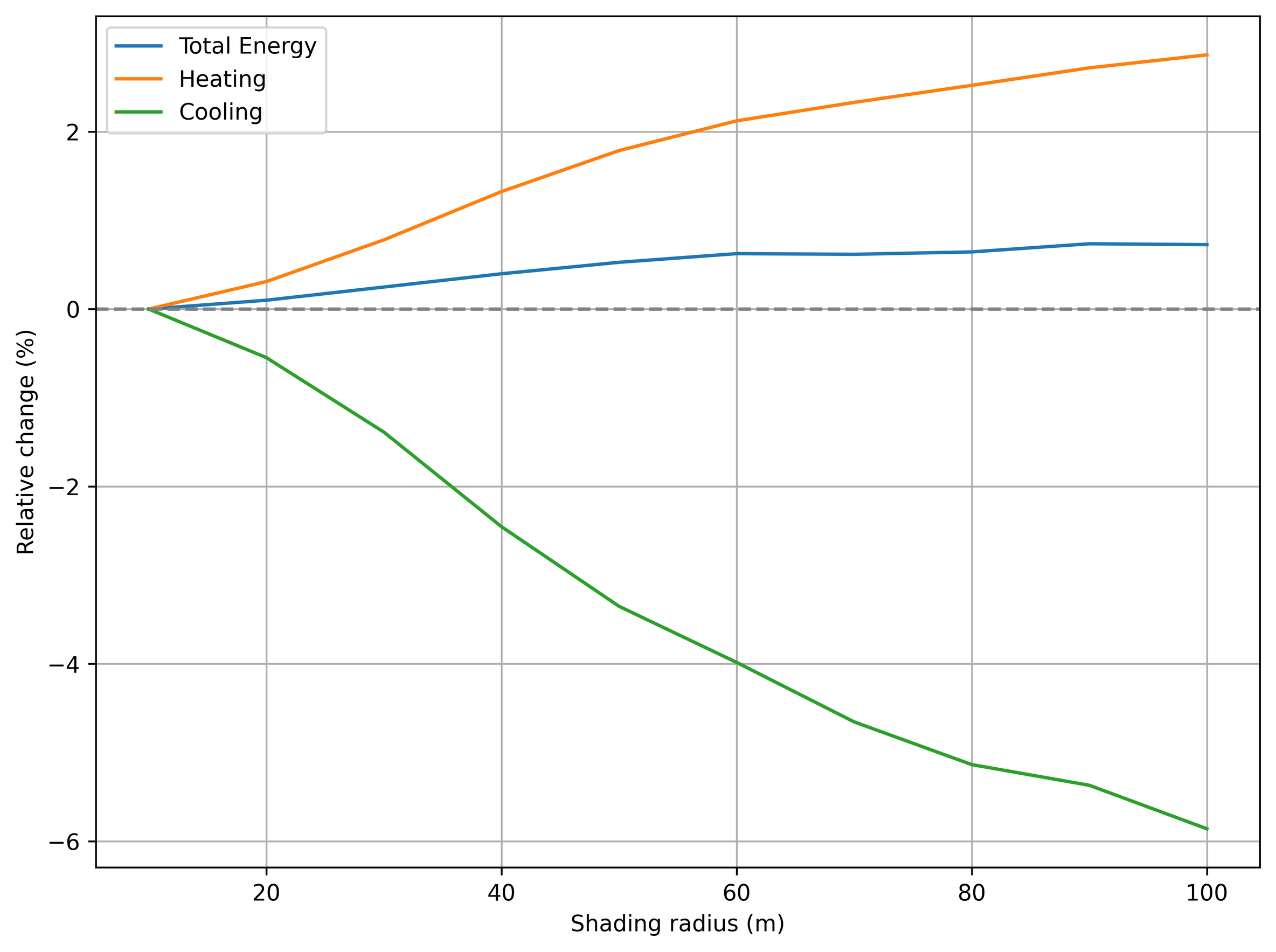}
	\caption{Relative change in mean energy intensity ($kWh/m^2$) across different shading radii.\label{figA.2.1}}
\end{figure}

The plot in Figure~\ref{figA.2.1} displays the percentage change in average energy intensity ($kWh/m^2$) across all buildings as a function of the shading buffer radius.
On the x-axis are the various shading buffers considered in the simulations.
For each radius, the corresponding value on the y-axis represents the relative change in mean energy intensity compared to the simulation with the minimum shading radius of 10m.
This valued is calculated using the following formula:
\begin{equation}
	\left(\frac{E_r}{E_{min}} - 1\right)\times 100
\end{equation}
Where:
\begin{itemize}
	\item $E_r$ is the mean energy intensity ($kWh/m\textsuperscript{2}$) at a given radius $r$.
	\item $E_{min}$ is the mean energy intensity ($kWh/m\textsuperscript{2}$) at the smallest radius of $10m$.
\end{itemize}

This normalization allows us to evaluate how sensitive the energy demand is to changes in shading radius, independent of absolute values.

\unskip

The results show that as the shading radius increases:
\begin{itemize}
	\item Heating demand increases by up to +3\%, due to reduced solar heat gains caused by the additional shading from surrounding buildings.
	      This indicates that larger buffers limit passive solar heating, especially in winter.
	\item Cooling demand decreases by up to - 6\%, as increased shading helps mitigate solar overheating during warmer months.
	\item The overall energy intensity change is positive, reflecting the net effect of increased heating demand outweighing the cooling energy savings in the simulated climate.
\end{itemize}

\begin{figure}[H]
	%\isPreprints{\centering}{} % Only used for preprints
	\centering
	\includegraphics[width=0.8\textwidth, keepaspectratio]{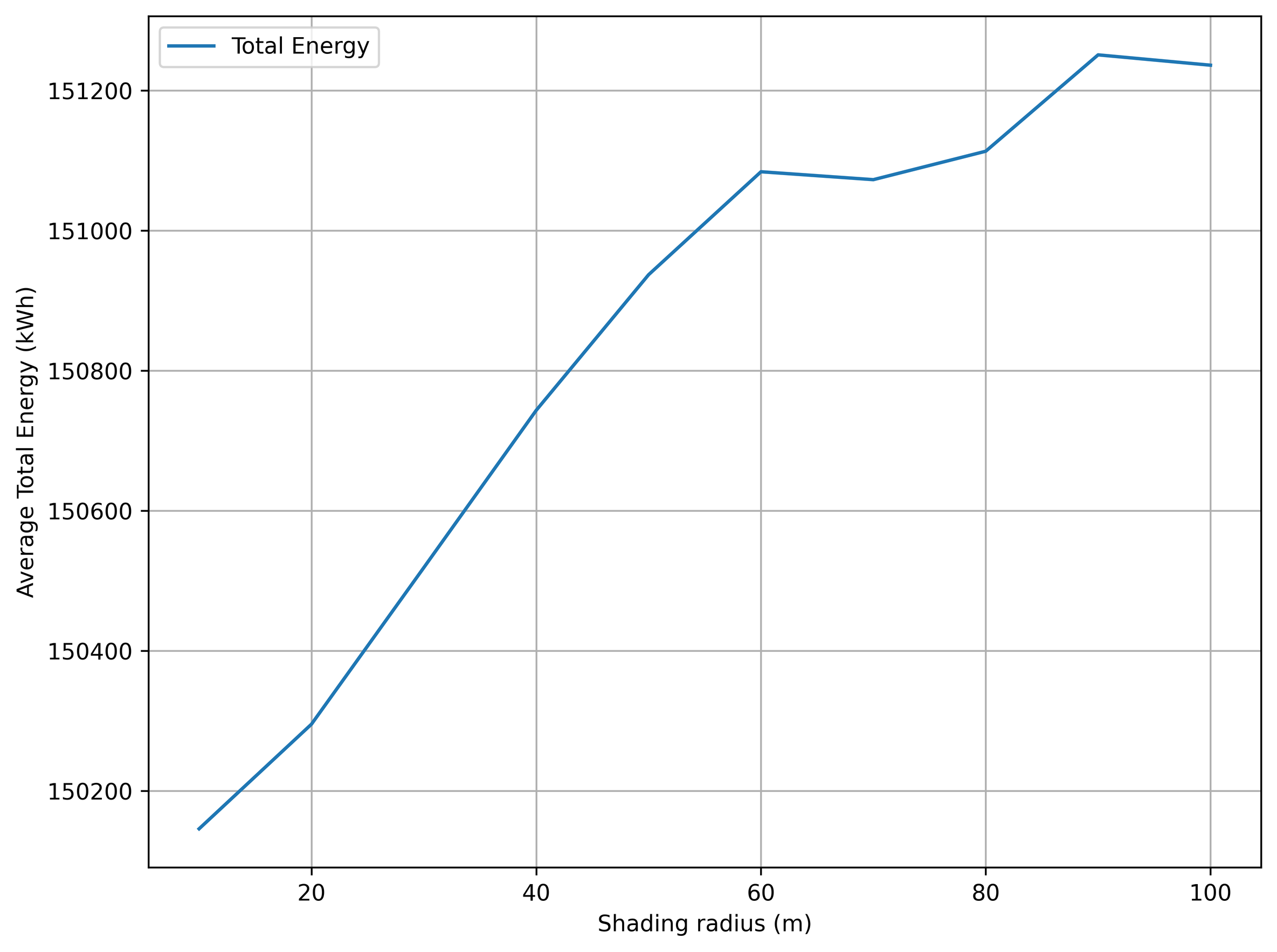}
	\caption{Average total energy as a function of the shading radius.\label{figA.2.2}}
\end{figure}

However, the percentage differences across radii are relatively small (e.g., within $\pm5 \- 6\%$).
Therefore, in addition to the considerations in the previous section, a mid-range radius provides a good balance between simulation accuracy and computational cost.
Moreover, if we specifically analyze the average total energy as a function of the shading radius, as shown in Figure~\ref{figA.2.2}, we can observe an  ``elbow`` around a radius of 60m.
Overall, the total energy demand exhibits only moderate variation across different radii.

\unskip

\section{Urban energy profile assessment:  analysis of simulated energy consumption for the current building stock} \label{current_energy_consumption}
\setcounter{figure}{0}
\renewcommand{\thefigure}{B\arabic{figure}}
This section presents the results obtained by simulating the energy consumption of the existing building stock.
To better understand the diversity in building energy consumption, we first explore the distribution of total energy use per square meter ($kWh/m^2$) grouped by the archetypes.
This allows us to assess how energy needs vary with building characteristics.

\begin{figure}[H]
	%\isPreprints{\centering}{} % Only used for preprints
	\centering
	\includegraphics[width=0.8\textwidth, keepaspectratio]{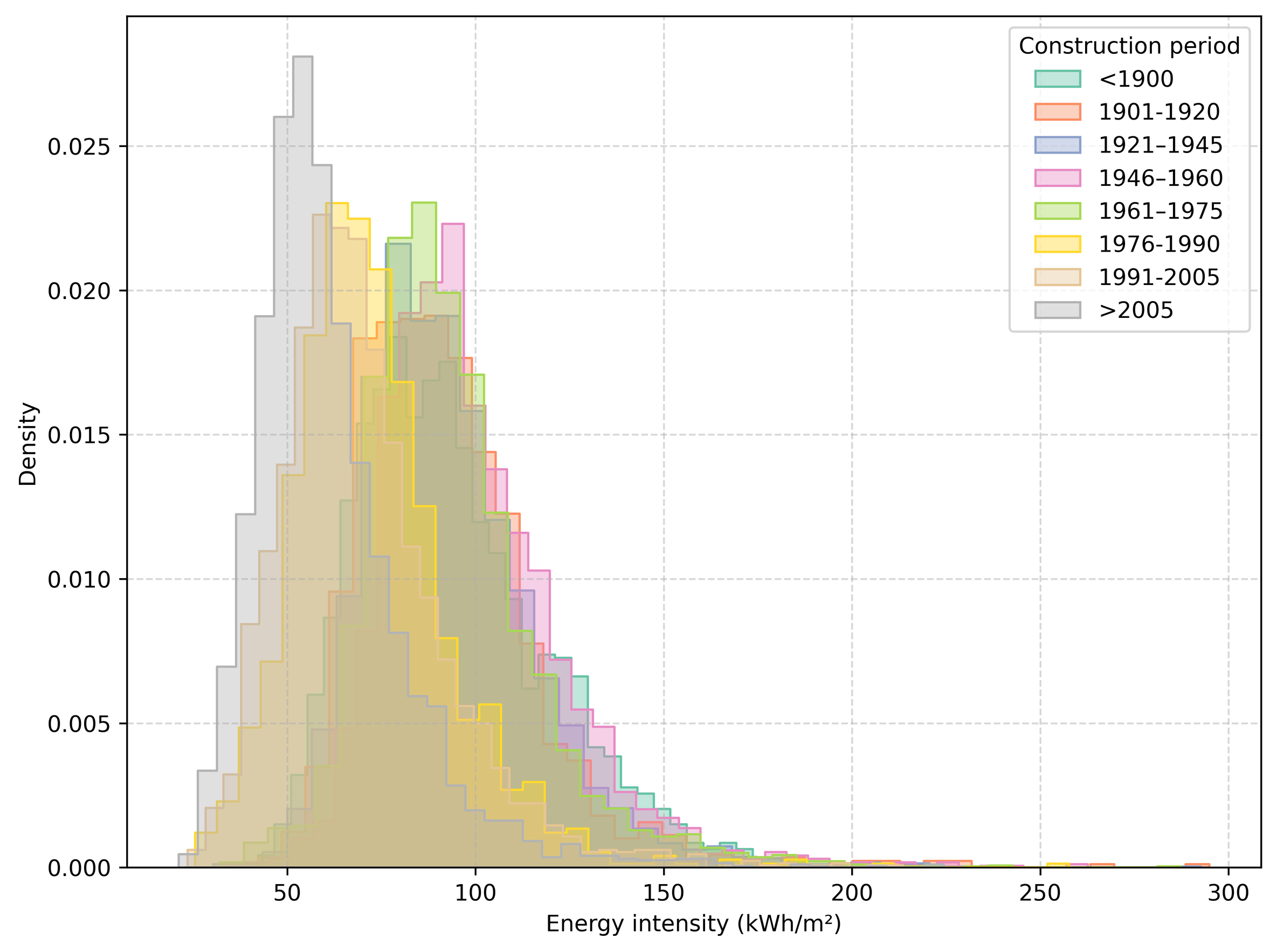}
	\caption{Energy intensity distribution by period of construction.\label{figC.1.1}}
\end{figure}

As shown in Figure~\ref{figC.1.1}, the histograms reveal differences in energy intensity ($kWh/m^2$) across buildings from different construction periods.
Older buildings tend to show broader and right-skewed distributions, indicating a higher variability and, generally, greater energy demands.
This is likely due to poorer thermal performance, limited insulation, and outdated envelope standards typical of early 20th-century construction.
In contrast, more recent buildings, particularly those built after 1975, show narrower and left-shifted distributions, reflecting improved energy performance likely due to stricter building codes and more efficient construction techniques.
The progressive shift toward lower energy intensity over time highlights the impact of energy regulations and retrofitting practices on reducing energy demand per square meter.

\unskip

Pareto analysis helps to identify the most significant factors contributing to an outcome, allowing for prioritized actions where they matter most.
In the context of building energy consumption, Pareto analysis is often used to identify which buildings or zones consume the most energy, allowing retrofits and policy interventions to be focused on high-impact areas.
Typically, about 80\% of the effects stem from 20\% of the causes. While our situation yields different results, they can still be useful for targeted action.

\begin{figure}[H]
	%\isPreprints{\centering}{} % Only used for preprints
	\centering
	\includegraphics[width=0.8\textwidth, keepaspectratio]{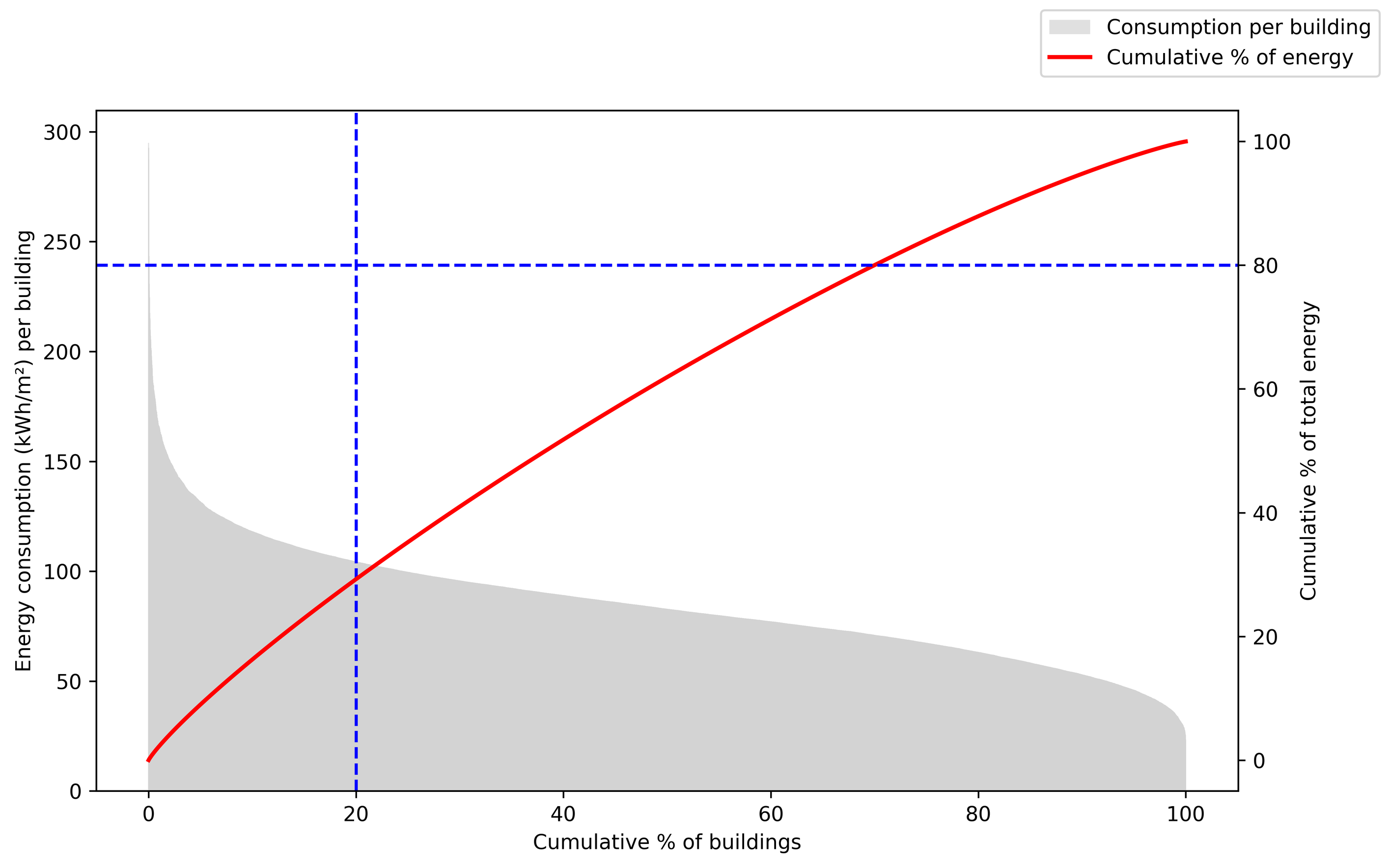}
	\caption{Pareto cumulative energy intensity consumption.\label{figC.1.2}}
\end{figure}

The Pareto plot in Figure~\ref{figC.1.2} reveals that  almost 70\% of buildings account for 80\% of the total energy consumption in terms of total energy consumption.
This demonstrates a distribution flatter than expected in typical urban energy datasets and suggests that high energy consumption is not highly concentrated in a small subset of buildings, but rather widely distributed across much of the stock.
As a result, energy efficiency policies must be broadly targeted, instead of focusing only on a few outliers.

\unskip

\begin{figure}[H]
	%\isPreprints{\centering}{} % Only used for preprints
	\centering
	\includegraphics[width=0.8\textwidth, keepaspectratio]{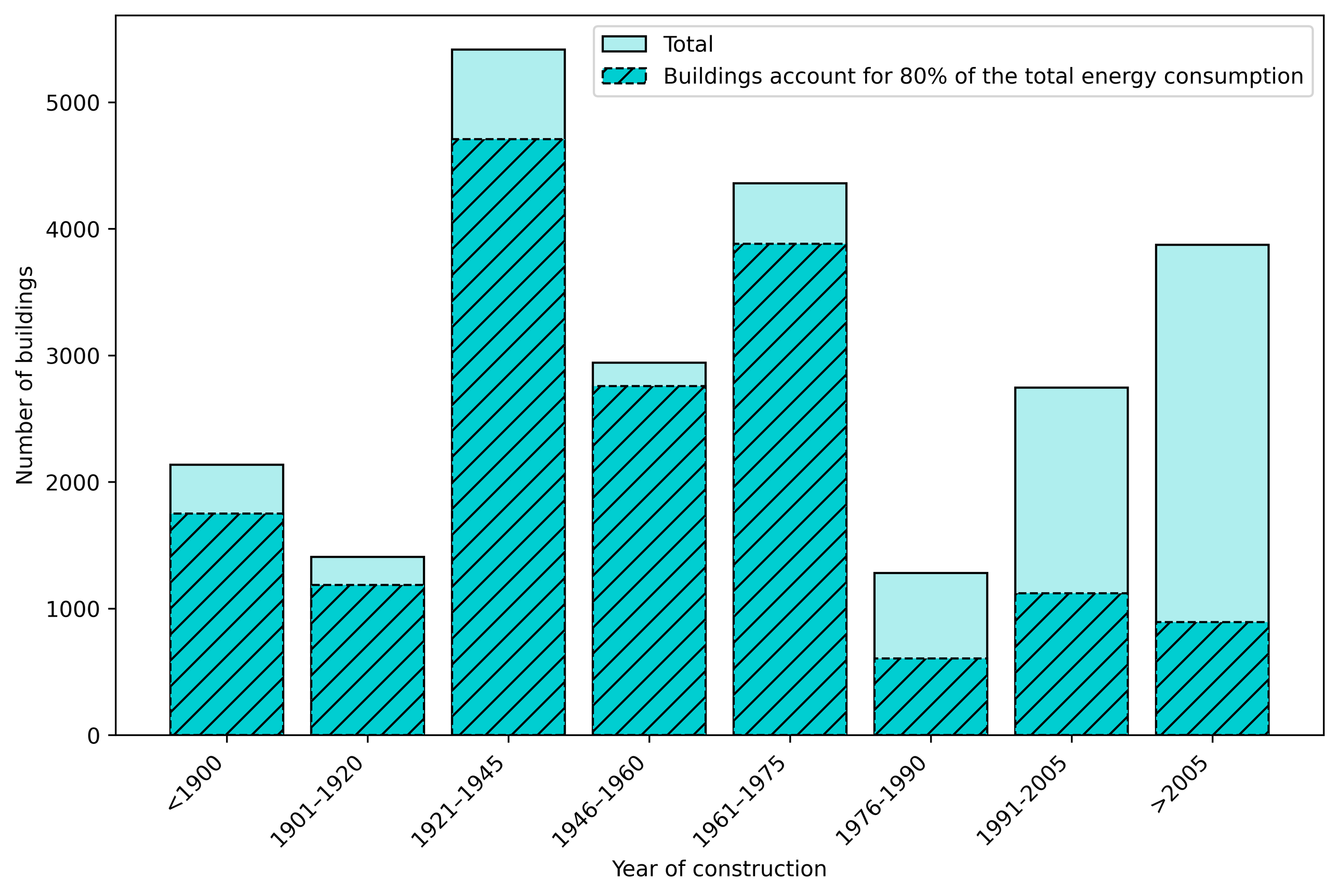}
	\caption{Year distribution of most consuming buildings.\label{figC.1.3}}
\end{figure}

Figure~\ref{figC.1.3} shows that most of the buildings with the highest energy consumption obtained with the Pareto analysis are the oldest ones or were built in the middle of the past century, when construction laws were less restrictive in terms of efficiency.

\unskip

After simulating the building's energy consumption, it is possible to estimate its $CO_2$ emissions. To do so, we can
compute an
approximation using the emission factor, which is a conversion factor that depends on the fuel used to generate energy.
This $CO_2$ estimate doesn’t consider other greenhouse gases; therefore, we cannot talk about the equivalent $CO_2$,
but it’s a good approximation. We used the emission factor provided by ISPRA in document \cite{ISPRA2025NID} for the year 2023,
which is $59.182 tCO_2/TJ$ and obtained an average $CO_2$ emission of 12.2 $kg/m^2$. This value likely underestimates the
actual value because it only considers $CO_2$ emissions from gas consumption, not from electricity or other greenhouse
gases.

%%%%%%%%%%%%%%%%%%%%%%%%%%%%%%%%%%%%%%%%%%
%\isPreprints{}{% This command is only used for ``preprints''.
\begin{adjustwidth}{-\extralength}{0cm}
	%} % If the paper is ``preprints'', please uncomment this parenthesis.
	%\printendnotes[custom] % Un-comment to print a list of endnotes
\end{adjustwidth}

\newpage
\begin{adjustwidth}{-\extralength}{0cm}
	\bibliography{refs}

@misc{buildings_IEA,
  author       = {International Energy Agency},
  howpublished = {\url{https://www.iea.org/energy-system/buildings}}
}

@article{Ferrando_2020,
title={Urban building energy modeling (UBEM) tools: A state-of-the-art review of bottom-up physics-based approaches},
volume={62},
ISSN={2210-6707},
url={http://dx.doi.org/10.1016/j.scs.2020.102408},
DOI={10.1016/j.scs.2020.102408},
journal={Sustainable Cities and Society},
publisher={Elsevier BV},
author={Ferrando, Martina and Causone, Francesco and Hong, Tianzhen and Chen, Yixing},
year={2020},
month=nov, pages={102408} }

@article{CEREZODAVILA2016237,
title = {Modeling Boston: A workflow for the efficient generation and maintenance of urban building energy models from existing geospatial datasets},
journal = {Energy},
volume = {117},
pages = {237-250},
year = {2016},
issn = {0360-5442},
doi = {https://doi.org/10.1016/j.energy.2016.10.057},
url = {https://www.sciencedirect.com/science/article/pii/S0360544216314918},
author = {Carlos {Cerezo Davila} and Christoph F. Reinhart and Jamie L. Bemis}
}

@article{REINHART2016196,
title = {Urban building energy modeling – A review of a nascent field},
journal = {Building and Environment},
volume = {97},
pages = {196-202},
year = {2016},
issn = {0360-1323},
doi = {https://doi.org/10.1016/j.buildenv.2015.12.001},
url = {https://www.sciencedirect.com/science/article/pii/S0360132315003248},
author = {Christoph F. Reinhart and Carlos {Cerezo Davila}}
}

@article{JOHARI2020109902,
title = {Urban building energy modeling: State of the art and future prospects},
journal = {Renewable and Sustainable Energy Reviews},
volume = {128},
pages = {109902},
year = {2020},
issn = {1364-0321},
doi = {https://doi.org/10.1016/j.rser.2020.109902},
url = {https://www.sciencedirect.com/science/article/pii/S1364032120301933},
author = {F. Johari and G. Peronato and P. Sadeghian and X. Zhao and J. Widén}
}

@article{ABBASABADI2019106270,
title = {Urban energy use modeling methods and tools: A review and an outlook},
journal = {Building and Environment},
volume = {161},
pages = {106270},
year = {2019},
issn = {0360-1323},
doi = {https://doi.org/10.1016/j.buildenv.2019.106270},
url = {https://www.sciencedirect.com/science/article/pii/S0360132319304809},
author = {Narjes Abbasabadi and Mehdi Ashayeri}
}

@misc{EnergyPlus940,
  author       = {{U.S. Department of Energy}},
  title        = {EnergyPlus, Version 9.4.0},
  howpublished = {\url{https://energyplus.net}},
  note         = {National Renewable Energy Laboratory}
}

@inproceedings{UMI,
author = {Reinhart, C. and Dogan, Timur and Jakubiec, J. and Rakha, Tarek and Sang, Andrew},
year = {2013},
month = {08},
pages = {},
title = {Umi – An Urban Simulation Environment For Building Energy Use, Daylighting And Walkability},
journal = {Proceedings of BS 2013: 13th Conference of the International Building Performance Simulation Association},
doi = {10.26868/25222708.2013.1404}
}

@article{tabula,
author = {Corrado, Vincenzo and Ballarini, Ilaria and Corgnati, Stefano and Talà, Novella},
year = {2011},
month = {01},
pages = {1-117},
title = {Building typology brochure-Italy fascicolo sulla tipologia edilizia italiana-Tabula eU project},
journal = {Politecnico di Torino, Torino}
}

@article{CORRADO2016200,
title = {Verification of the New Ministerial Decree about Minimum Requirements for the Energy Performance of Buildings},
journal = {Energy Procedia},
volume = {101},
pages = {200-207},
year = {2016},
note = {ATI 2016 - 71st Conference of the Italian Thermal Machines Engineering Association},
issn = {1876-6102},
doi = {https://doi.org/10.1016/j.egypro.2016.11.026},
url = {https://www.sciencedirect.com/science/article/pii/S1876610216312358},
author = {Vincenzo Corrado and Ilaria Ballarini and Domenico Dirutigliano and Giovanni Murano}
}

@article{CORRADO2018,
author = {Ballarini, Ilaria and Primo, Elisa and Corrado, Vincenzo},
year = {2018},
month = {01},
pages = {133-133},
title = {On the limits of the quasi-steady-state method to predict the energy performance of low-energy buildings},
volume = {2018},
journal = {Thermal Science},
doi = {10.2298/TSCI170724133B}
}

@techreport{ISPRA2025NID,
  author       = {Daniela Romano and Antonella Bernetti and Antonio Caputo and Marco Cordella 
                  and Riccardo De Lauretis and Eleonora Di Cristofaro and Angela Fiore 
                  and Andrea Gagna and Barbara Gonella and Federica Moricci and Guido Pellis 
                  and Ernesto Taurino and Marina Vitullo},
  title        = {Italian Greenhouse Gas Inventory 1990--2023. National Inventory Document 2025},
  institution  = {ISPRA -- Istituto Superiore per la Protezione e la Ricerca Ambientale},
  year         = {2025},
  number       = {Rapporti 411/2025},
  address      = {Rome, Italy},
  isbn         = {978-88-448-1252-2},
  url          = {http://www.isprambiente.gov.it/},
  note         = {Annual Report for submission under the UNFCCC and the Paris Agreement}
}

@article{KONTOKOSTA2017303,
title = {A data-driven predictive model of city-scale energy use in buildings},
journal = {Applied Energy},
volume = {197},
pages = {303-317},
year = {2017},
issn = {0306-2619},
doi = {https://doi.org/10.1016/j.apenergy.2017.04.005},
url = {https://www.sciencedirect.com/science/article/pii/S0306261917303835},
author = {Constantine E. Kontokosta and Christopher Tull}
}

@article{CHEN2017323,
title = {Automatic generation and simulation of urban building energy models based on city datasets for city-scale building retrofit analysis},
journal = {Applied Energy},
volume = {205},
pages = {323-335},
year = {2017},
issn = {0306-2619},
doi = {https://doi.org/10.1016/j.apenergy.2017.07.128},
url = {https://www.sciencedirect.com/science/article/pii/S0306261917310024},
author = {Yixing Chen and Tianzhen Hong and Mary Ann Piette}
}

@article{CHEN2020115584,
title = {Automatic and rapid calibration of urban building energy models by learning from energy performance database},
journal = {Applied Energy},
volume = {277},
pages = {115584},
year = {2020},
issn = {0306-2619},
doi = {https://doi.org/10.1016/j.apenergy.2020.115584},
url = {https://www.sciencedirect.com/science/article/pii/S0306261920310953},
author = {Yixing Chen and Zhang Deng and Tianzhen Hong}
}

@article{FONSECA_2016,
title = {City Energy Analyst (CEA): Integrated framework for analysis and optimization of building energy systems in neighborhoods and city districts},
journal = {Energy and Buildings},
volume = {113},
pages = {202-226},
year = {2016},
issn = {0378-7788},
doi = {https://doi.org/10.1016/j.enbuild.2015.11.055},
url = {https://www.sciencedirect.com/science/article/pii/S0378778815304199},
author = {Jimeno A. Fonseca and Thuy-An Nguyen and Arno Schlueter and Francois Marechal}
}

@inproceedings{hong_2016,
author = {Hong, Tianzhen and Chen, Yixing and Lee, Sang Hoon and Piette, M.},
year = {2016},
month = {08},
pages = {},
title = {CityBES: A Web-based Platform to Support City-Scale Building Energy Efficiency}
}

@Article{urbansci9040113,
AUTHOR = {Espino-Reyes, Carlos A. and Ortega-Avila, Naghelli and Lucero-Álvarez, Jorge and Rodríguez-Muñoz, Norma A.},
TITLE = {Energy Consumption in Mexican Homes: Using a Reference Building as a Launchpad for Achieving Nearly Zero Energy},
JOURNAL = {Urban Science},
VOLUME = {9},
YEAR = {2025},
NUMBER = {4},
ARTICLE-NUMBER = {113},
URL = {https://www.mdpi.com/2413-8851/9/4/113},
ISSN = {2413-8851},
DOI = {10.3390/urbansci9040113}
}

@Article{urbansci7030096,
AUTHOR = {Niza, Iasmin Lourenço and Bueno, Ana Maria and Broday, Evandro Eduardo},
TITLE = {Indoor Environmental Quality (IEQ) and Sustainable Development Goals (SDGs): Technological Advances, Impacts and Challenges in the Management of Healthy and Sustainable Environments},
JOURNAL = {Urban Science},
VOLUME = {7},
YEAR = {2023},
NUMBER = {3},
ARTICLE-NUMBER = {96},
URL = {https://www.mdpi.com/2413-8851/7/3/96},
ISSN = {2413-8851},
DOI = {10.3390/urbansci7030096}
}

@article{MALHOTRA2022108552,
title = {Information modelling for urban building energy simulation—A taxonomic review},
journal = {Building and Environment},
volume = {208},
pages = {108552},
year = {2022},
issn = {0360-1323},
doi = {https://doi.org/10.1016/j.buildenv.2021.108552},
url = {https://www.sciencedirect.com/science/article/pii/S0360132321009422},
author = {Avichal Malhotra and Julian Bischof and Alexandru Nichersu and Karl-Heinz Häfele and Johannes Exenberger and Divyanshu Sood and James Allan and Jérôme Frisch and Christoph {van Treeck} and James O’Donnell and Gerald Schweiger}
}

@Article{urbansci9100421,
AUTHOR = {Romeo, Taqir Mahmood and Ahsan, Tahmina and Zaman, Atiq},
TITLE = {Net-Zero Energy Retrofitting in Perth’s Residential Sector: Key Features and Strategies for Sustainable Building Transformation},
JOURNAL = {Urban Science},
VOLUME = {9},
YEAR = {2025},
NUMBER = {10},
ARTICLE-NUMBER = {421},
URL = {https://www.mdpi.com/2413-8851/9/10/421},
ISSN = {2413-8851},
DOI = {10.3390/urbansci9100421}
}

@article{ORAIOPOULOS_2022,
title = {On the accuracy of Urban Building Energy Modelling},
journal = {Renewable and Sustainable Energy Reviews},
volume = {158},
pages = {111976},
year = {2022},
issn = {1364-0321},
doi = {https://doi.org/10.1016/j.rser.2021.111976},
url = {https://www.sciencedirect.com/science/article/pii/S1364032121012405},
author = {A. Oraiopoulos and B. Howard}
}

@article{ALI2021111073,
title = {Review of urban building energy modeling (UBEM) approaches, methods and tools using qualitative and quantitative analysis},
journal = {Energy and Buildings},
volume = {246},
pages = {111073},
year = {2021},
issn = {0378-7788},
doi = {https://doi.org/10.1016/j.enbuild.2021.111073},
url = {https://www.sciencedirect.com/science/article/pii/S0378778821003571},
author = {Usman Ali and Mohammad Haris Shamsi and Cathal Hoare and Eleni Mangina and James O’Donnell}
}

@article{HONG2020106508,
title = {Ten questions on urban building energy modeling},
journal = {Building and Environment},
volume = {168},
pages = {106508},
year = {2020},
issn = {0360-1323},
doi = {https://doi.org/10.1016/j.buildenv.2019.106508},
url = {https://www.sciencedirect.com/science/article/pii/S0360132319307206},
author = {Tianzhen Hong and Yixing Chen and Xuan Luo and Na Luo and Sang Hoon Lee}
}

@article{FERRANDO2020102408,
title = {Urban building energy modeling (UBEM) tools: A state-of-the-art review of bottom-up physics-based approaches},
journal = {Sustainable Cities and Society},
volume = {62},
pages = {102408},
year = {2020},
issn = {2210-6707},
doi = {https://doi.org/10.1016/j.scs.2020.102408},
url = {https://www.sciencedirect.com/science/article/pii/S2210670720306296},
author = {Martina Ferrando and Francesco Causone and Tianzhen Hong and Yixing Chen}
}

@article{JOHARI_2020,
title = {Urban building energy modeling: State of the art and future prospects},
journal = {Renewable and Sustainable Energy Reviews},
volume = {128},
pages = {109902},
year = {2020},
issn = {1364-0321},
doi = {https://doi.org/10.1016/j.rser.2020.109902},
url = {https://www.sciencedirect.com/science/article/pii/S1364032120301933},
author = {F. Johari and G. Peronato and P. Sadeghian and X. Zhao and J. Widén}
}

@article{DENG_2023,
title = {Using urban building energy modeling to quantify the energy performance of residential buildings under climate change},
journal = {Building Simulation},
year = {2023},
issn = {1996-3599},
doi = {https://doi.org/10.1007/s12273-023-1032-2},
url = {https://link.springer.com/article/10.1007/s12273-023-1032-2},
author = {Z. Deng and K. Javanroodi and V. M. Nik and Y. Chen}
}

@Article{su13041595,
AUTHOR = {Todeschi, Valeria and Boghetti, Roberto and Kämpf, Jérôme H. and Mutani, Guglielmina},
TITLE = {Evaluation of Urban-Scale Building Energy-Use Models and Tools—Application for the City of Fribourg, Switzerland},
JOURNAL = {Sustainability},
VOLUME = {13},
YEAR = {2021},
NUMBER = {4},
ARTICLE-NUMBER = {1595},
URL = {https://www.mdpi.com/2071-1050/13/4/1595},
ISSN = {2071-1050},
DOI = {10.3390/su13041595}
}

@article{WEI20181027,
title = {A review of data-driven approaches for prediction and classification of building energy consumption},
journal = {Renewable and Sustainable Energy Reviews},
volume = {82},
pages = {1027-1047},
year = {2018},
issn = {1364-0321},
doi = {https://doi.org/10.1016/j.rser.2017.09.108},
url = {https://www.sciencedirect.com/science/article/pii/S136403211731362X},
author = {Yixuan Wei and Xingxing Zhang and Yong Shi and Liang Xia and Song Pan and Jinshun Wu and Mengjie Han and Xiaoyun Zhao}
}

@article{CEREZO2017321,
title = {Comparison of four building archetype characterization methods in urban building energy modeling (UBEM): A residential case study in Kuwait City},
journal = {Energy and Buildings},
volume = {154},
pages = {321-334},
year = {2017},
issn = {0378-7788},
doi = {https://doi.org/10.1016/j.enbuild.2017.08.029},
url = {https://www.sciencedirect.com/science/article/pii/S0378778817314743},
author = {Carlos Cerezo and Julia Sokol and Saud AlKhaled and Christoph Reinhart and Adil Al-Mumin and Ali Hajiah}
}

@Article{en13164244,
AUTHOR = {Goy, Solène and Maréchal, François and Finn, Donal},
TITLE = {Data for Urban Scale Building Energy Modelling: Assessing Impacts and Overcoming Availability Challenges},
JOURNAL = {Energies},
VOLUME = {13},
YEAR = {2020},
NUMBER = {16},
ARTICLE-NUMBER = {4244},
URL = {https://www.mdpi.com/1996-1073/13/16/4244},
ISSN = {1996-1073},
DOI = {10.3390/en13164244}
}

@article{BLANCO2024105075,
title = {Data-driven classification of Urban Energy Units for district-level heating and electricity demand analysis},
journal = {Sustainable Cities and Society},
volume = {101},
pages = {105075},
year = {2024},
issn = {2210-6707},
doi = {https://doi.org/10.1016/j.scs.2023.105075},
url = {https://www.sciencedirect.com/science/article/pii/S2210670723006856},
author = {Luis Blanco and Alaa Alhamwi and Björn Schiricke and Bernhard Hoffschmidt}
}

@Article{su12145678,
AUTHOR = {Mutani, Guglielmina and Todeschi, Valeria and Beltramino, Simone},
TITLE = {Energy Consumption Models at Urban Scale to Measure Energy Resilience},
JOURNAL = {Sustainability},
VOLUME = {12},
YEAR = {2020},
NUMBER = {14},
ARTICLE-NUMBER = {5678},
URL = {https://www.mdpi.com/2071-1050/12/14/5678},
ISSN = {2071-1050},
DOI = {10.3390/su12145678}
}

@article{Ji_2022,
author = {Ji, Qunfeng and Yangbo, Bi and Makvandi, Mehdi and Xilin, Zhou and Chuancheng, Li},
year = {2022},
month = {03},
pages = {},
title = {Modelling Building Stock Energy Consumption at the Urban Level from an Empirical Study},
journal = {Buildings},
doi = {10.3390/buildings12030385}
}

@Article{urbansci9110439,
AUTHOR = {Garcia-Nevado, Elena and Lopez-Besora, Judit and Besuievsky, Gonzalo},
TITLE = {Including Open Balconies in Housing Retrofitting: A Parametric Analysis for Energy Efficiency},
JOURNAL = {Urban Science},
VOLUME = {9},
YEAR = {2025},
NUMBER = {11},
ARTICLE-NUMBER = {439},
URL = {https://www.mdpi.com/2413-8851/9/11/439},
ISSN = {2413-8851},
DOI = {10.3390/urbansci9110439}
}

@Article{urbansci8040215,
AUTHOR = {Rodrigues, André and Oliveira, Armando C. and Palmero-Marrero, Ana I.},
TITLE = {Integration of PV Systems into the Urban Environment: A Review of Their Effects and Energy Models},
JOURNAL = {Urban Science},
VOLUME = {8},
YEAR = {2024},
NUMBER = {4},
ARTICLE-NUMBER = {215},
URL = {https://www.mdpi.com/2413-8851/8/4/215},
ISSN = {2413-8851},
DOI = {10.3390/urbansci8040215}
}

@Article{urbansci_khan_2025,
AUTHOR = {Khan, Kaleem Ullah and Ali, Ghaffar and Murtaza, Natasha and Pan, Yanchun and Kysucky, Vlado},
TITLE = {Toward Net-Zero Emissions: The Role of Smart City Technologies in Reducing Carbon Emissions in China},
JOURNAL = {Urban Science},
VOLUME = {9},
YEAR = {2025},
NUMBER = {9},
ARTICLE-NUMBER = {374},
URL = {https://www.mdpi.com/2413-8851/9/9/374},
ISSN = {2413-8851},
DOI = {10.3390/urbansci9090374}
}

@Article{urbansci_kasmeridis_2025,
AUTHOR = {Kasmeridis, Ilias K. and Skandalos, Nikolaos and Dimitriou, Tsampika and Dimakopoulos, Vassilios V. and Karamanis, Dimitrios},
TITLE = {Digitized Energy Systems and Open-Access Platforms: Accelerating Cities’ Transition to Carbon Neutrality},
JOURNAL = {Urban Science},
VOLUME = {9},
YEAR = {2025},
NUMBER = {9},
ARTICLE-NUMBER = {364},
URL = {https://www.mdpi.com/2413-8851/9/9/364},
ISSN = {2413-8851},
DOI = {10.3390/urbansci9090364}
}

@article{BUCKLEY_2021,
title = {Using urban building energy modelling (UBEM) to support the new European Union’s Green Deal: Case study of Dublin Ireland},
journal = {Energy and Buildings},
volume = {247},
pages = {111115},
year = {2021},
issn = {0378-7788},
doi = {https://doi.org/10.1016/j.enbuild.2021.111115},
url = {https://www.sciencedirect.com/science/article/pii/S0378778821003996},
author = {Niall Buckley and Gerald Mills and Christoph Reinhart and Zachary Michael Berzolla}
}

@article{FOUCQUIER_2013,
title = {State of the art in building modelling and energy performances prediction: A review},
journal = {Renewable and Sustainable Energy Reviews},
volume = {23},
pages = {272-288},
year = {2013},
issn = {1364-0321},
doi = {https://doi.org/10.1016/j.rser.2013.03.004},
url = {https://www.sciencedirect.com/science/article/pii/S1364032113001536},
author = {Aurélie Foucquier and Sylvain Robert and Frédéric Suard and Louis Stéphan and Arnaud Jay}
}
\end{adjustwidth}

\end{document}